\documentclass[preprint]{sigplanconf}

\usepackage{epsfig}
\usepackage{color}
\usepackage{amsmath}
\usepackage{amsfonts}
\usepackage{amssymb} 
\usepackage{amsthm}
\usepackage{comment}
\usepackage{multirow}
\usepackage{algorithm}
\usepackage{algorithmic}

\begin{document}

\setlength{\pdfpageheight}{\paperheight}
\setlength{\pdfpagewidth}{\paperwidth}

\conferenceinfo{CONF 'yy}{Month d--d, 20yy, City, ST, Country} 
\copyrightyear{20yy} 
\copyrightdata{978-1-nnnn-nnnn-n/yy/mm} 
\doi{nnnnnnn.nnnnnnn}


\title{On data skewness, stragglers, and MapReduce\\progress indicators}


\authorinfo{Emilio Coppa\and Irene Finocchi}
           {Computer Science Department, Sapienza University of Rome}
           {{\tt\{coppa, finocchi\}@di.uniroma1.it}}

\maketitle

\begin{abstract}
We tackle the problem of predicting the performance of MapReduce applications designing accurate {\em progress indicators},  which keep programmers informed on the percentage of completed computation time during the execution of a job. Through extensive experiments, we show that state-of-the-art progress indicators (including the one provided by Hadoop) can be seriously harmed by data skewness, load unbalancing, and straggling tasks. This is mainly due  to their implicit assumption that the running time depends linearly on the input size. We thus design a novel profile-guided progress indicator, called {\tt NearestFit}, that operates without the linear hypothesis assumption and exploits a careful combination of nearest neighbor regression and statistical curve fitting techniques. 
Our theoretical progress model requires fine-grained profile data, that can be very difficult to manage in practice. To overcome this issue, we resort to computing accurate approximations for some of the quantities used in our model through space- and time-efficient data streaming algorithms. 
We implemented {\tt NearestFit} on top of Hadoop 2.6.0. An extensive empirical assessment over the Amazon EC2 platform on a variety of real-world benchmarks shows that {\tt NearestFit}  is practical w.r.t. space and time overheads and that its accuracy is generally very good, even in scenarios where competitors incur non-negligible errors and wide prediction fluctuations.
Overall, {\tt NearestFit} significantly improves the current state-of-art on progress analysis for MapReduce.
\end{abstract}



\section{Introduction}
\label{se:intro}

The ability to perform scalable and timely processing of massive datasets is a crucial endeavor of our era. To handle increasing volumes of data, the last fifteen years have seen the emergence of powerful computational infrastructures that, in turn, have spurred new programming languages technologies, resulting in a generalized shift from sequential to parallel programming models. 

The quest for processing extreme data on complex platforms in a programmer-accessible way has been key to the success of MapReduce~\cite{DG08} and of the entire Apache ecosystem centered around Hadoop~\cite{hadoop}. MapReduce allows developers to expose parallelism in their applications by means of powerful high-level computing primitives (map and reduce functions), hiding  the details of how a computation is actually mapped to the underlying distributed platform. Its runtime system automatically parallelizes the computation, handling complex low-level details of the execution (data partitioning, task distribution, load balancing, node communication,  fault tolerance) and making it possible to scale applications to large clusters  of inexpensive commodity nodes. Since the introduction of MapReduce in 2004, there has been a proliferation of programming models and software frameworks for large-scale data analysis in response to diverse application requirements (e.g., iterative processing~\cite{twister10, spark12}, streaming~\cite{S4}, incremental computations~\cite{Incoop}, SQL-like languages~\cite{pig08, hive10}, and graph processing~\cite{pregel10}). 

Big data systems have significantly improved software development at scale: the ease of programming and the capability to express ample sets of algorithms have been primary concerns in their design. However,  these frameworks typically turn out to be a ``black box'' to programmers, who are still faced with many diverse and difficult issues when debugging and optimizing applications. For instance, when a user runs a MapReduce job that seems to take an abnormally long time, there is no easy way of pinpointing the reason for that behavior. Our own experience is confirmed by anecdotal evidence in many developers' forums, where programmers often indicate unexpected performance behaviors asking for insights.

\vspace{2mm}
\noindent{\bf Progress analysis.} An important problem targeted by a variety of works in the last few years is to predict the performance of MapReduce applications designing accurate {\em progress indicators} (see, e.g.,~\cite{parallax,paratimer,ARIA11,microbench}). A progress indicator keeps programmers informed on the percentage of completed computation time during the execution of a job. This is especially important for long-running applications and can shed light into abnormal behaviors, helping programmers distinguish between slow or stalled computations and pinpointing algorithmic inefficiencies, programming errors, or load balancing issues. In certain settings, especially in pay-as-you-go cloud enviroments, the user might want  to identify slow jobs and possibly abort them in order to avoid excessive costs. Progress analysis can also guide useful profile-driven optimizations, such as skew mitigation techniques~\cite{skewreduce,skewtune}. It is interesting to note that Hadoop~\cite{hadoop} comes with its own progress indicator.

A typical hypothesis in the design of progress indicators is that the running time depends linearly on the input size. For instance, the completion time of a task may be computed as the product between the size of the unprocessed data for that task and the average processing speed observed so far. This {\em linear progress assumption} can be a serious limitation in some applications, especially when the computational complexity of map/reduce functions grows more than linearly with respect to the input size. Such computations are not unfrequent. Computing the clustering coefficient, which is very useful in social network analysis, is one such example: state-of-the-art algorithms exploit reduce functions with quadratic worst-case running time~\cite{SV11}. Many recommendation systems for Web sites such as Yahoo! or LinkedIn are also more and more driven by computation-intensive analytics over massively distributed data. Large running times, combined with {\em data skewness} (e.g., power-law degree distributions in social networks) are responsible of the so-called {\em curse of the last reducer} phenomenon, where 99\% of the computation terminates quickly, but the remaining 1\% takes a disproportionately longer amount of time. Unfortunately, the wall-clock times of {\em stragglers} (slow running tasks) are far from being well-predicted under the linear progress assumption. 

Another widespread practice is to compute progress by exploiting profile data collected from previous executions. This can be sometimes misleading due to variability in platform and application parameters and to datasets with quite diverse characteristics: it may be easily the case that the same algorithm behaves very differently even on networks of rather similar size, depending on properties such as degree distribution or small-world phenomena~\cite{FFF14}. 

\begin{figure}[t]
\begin{center}
\includegraphics[width=0.99\columnwidth]{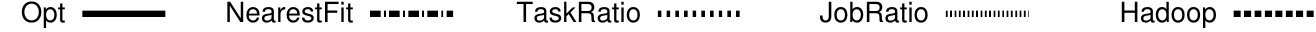}
\begin{tabular}{cc}
\hspace{-3mm}\includegraphics[width=0.50\columnwidth]{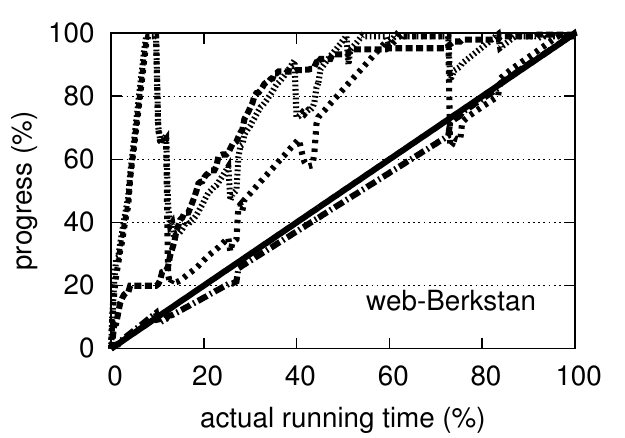} &
\hspace{-3mm}\includegraphics[width=0.50\columnwidth]{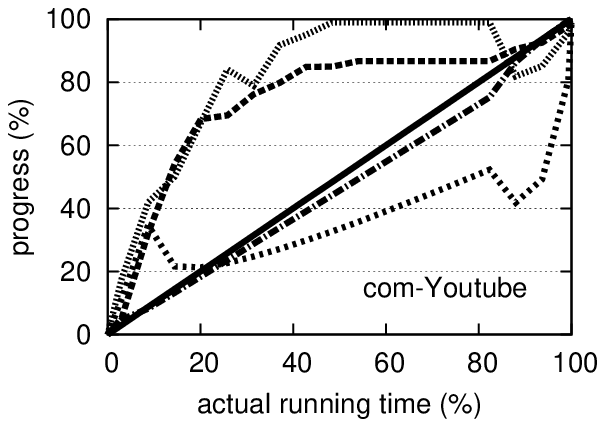} \\
\hspace{-3mm}\includegraphics[width=0.50\columnwidth]{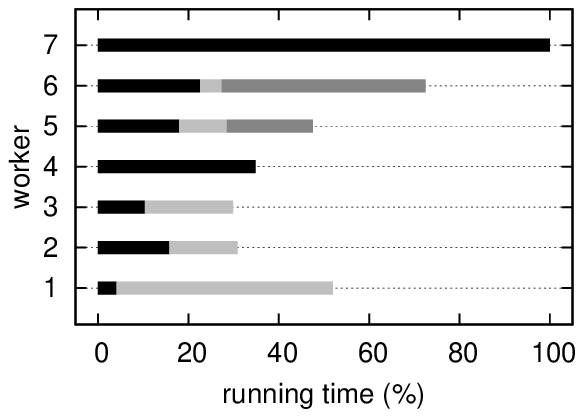} &
\hspace{-3mm}\includegraphics[width=0.50\columnwidth]{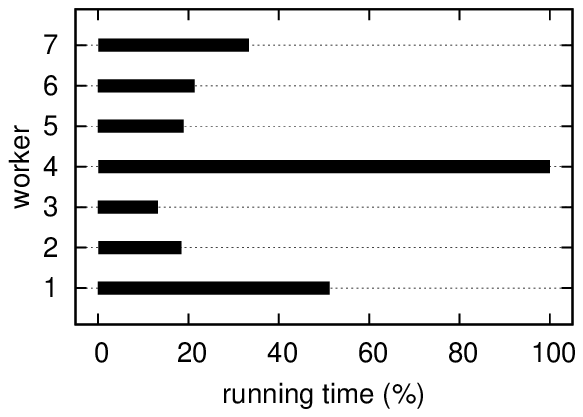} \\
\end{tabular}
\vspace{-3.5mm}
\end{center}
\caption{Computing clustering coefficients in networks {\tt web-} {\tt Berkstan} (left) and {\tt com-Youtube} (right). The upper charts plot progress estimates of different progress indicators. The lower charts (swimlanes plots) show load unbalancing among worker nodes.}
\label{fi:ni-bs-you}
\end{figure}

\vspace{2mm}
\noindent{\bf A motivating example.} Figure~\ref{fi:ni-bs-you} exemplifies some of the aforementioned issues. We used as a benchmark the {\tt NodeIterator} algorithm for computing clustering coefficients~\cite{SV11} applied to a web graph  and a social network from the SNAP project~\cite{snap}. The upper charts in Figure~\ref{fi:ni-bs-you} show the behavior of different progress indicators, reporting the actual progress on the $x$-axis vs. the estimated progress on the $y$-axis. State-of-the-art progress indicators (called {\tt Hadoop}, {\tt JobRatio}, and {\tt TaskRatio} in Figure~\ref{fi:ni-bs-you}) incur non-negligible errors and wide prediction fluctuations. For instance in {\tt com-Youtube}, after 4 minutes of execution (20\% of the actual running time) the default Hadoop progress indicator estimates that roughly 70\% of the computation is completed: the programmer will thus expect the execution to terminate in about 2 additional minutes, while the true wall-clock time will be considerably larger (18 minutes). Prediction errors are mainly due to stragglers, whose presence is shown in the swimlanes plots at the bottom of Figure~\ref{fi:ni-bs-you}. 

\vspace{3mm}
\noindent{\bf Our contribution.} Progress analysis issues exemplified by Figure~\ref{fi:ni-bs-you}  are the starting point for this paper. Our main contribution is the design and the implementation of a novel progress indicator that operates without the linear model assumption and exploits only dynamically collected profile data. Moving away from the linear hypothesis requires a complete rethinking of the underlying progress model, but offers significant benefits in the presence of  skewed data and computations with straggling tasks. Our progress indicator, called {\tt NearestFit}, significantly improves the current state-of-art on progress analysis. For instance, in Figure~\ref{fi:ni-bs-you} {\tt NearestFit} almost matches the optimal progress (straight line).
In more details, our main contributions can be summarized as follows:

\begin{itemize}

\item We formalize a {\em profile-guided theoretical progress model} based on a careful combination of {\em nearest neighbor regression} and statistical {\em curve fitting} (hence the name {\tt NearestFit}). While nearest neighbor regression is very stable over time and yields accurate estimates, resorting to curve fitting is beneficial to perform extrapolations, i.e., to predict the running times beyond the range of the already observed executions. This turns out to be especially useful in the presence of data skewness and straggling instances.

\item In order to apply regression analysis techniques, we profile the executions of map/reduce functions, collecting a set of data points that relate the running time of each invocation to its observed input size. Such {\em fine-grained profiles} are very difficult to manage in practice, yielding large time and space overheads. We nevertheless show that we can make {\tt NearestFit} practical: this is achieved through space-efficient data structures and {\em data streaming algorithms}, which enable the efficient computation of accurate approximations for some of the quantities used in our model.

\item We perform an extensive empirical assessment of {\tt NearestFit} on the Amazon EC2 platform
using a variety of real-world benchmarks, which expose different computational patterns of MapReduce applications. We compare {\tt NearestFit} against state-of-the-art progress indicators showing that:

\begin{itemize}
\item the accuracy of {\tt NearestFit} is generally very good, even in the presence of data skewness and load unbalancing, which can seriously harm competitors;
\item the orderly combination of nearest neighbor regression and curve fitting is crucial to obtain good progress estimates: none of the techniques alone achieves reasonable results;
\item space and time overheads can be kept small;
\item the use of space-efficient streaming data structures guarantees efficiency, without affecting accuracy. 
\end{itemize}

\end{itemize}



\section{Anatomy of a MapReduce job}
\label{se:mapreduce}

Introduced by Google in 2004, MapReduce has emerged as one of the most popular systems for batch processing of extreme datasets~\cite{DG08}. Its functional-style programming model -- where  programmers  just need to implement two distinct {\em map} and {\em reduce functions} -- is indeed simple and amenable to a variety of real-world tasks, while low-level issues are automatically and transparently handled by the runtime system. 

Map and reduce functions are executed inside {\em map tasks} and {\em reduce tasks}, respectively. Both of them work on data represented as $\langle key, value\rangle$ pairs. Input and output pairs are
stored on a distributed file system, such as Google's file system or HDFS.  

\medskip
\noindent {\bf Inner workings of MapReduce.} The runtime system splits the job input into fixed-size chunks (e.g., 64MB each), spawning a map task per chunk. Depending on cluster capacity (i.e., number of worker nodes), multiple map tasks can be run in parallel. Each map task scans its input chunk, invoking the map function on each $\langle k, v \rangle$ pair. A map invocation can emit a list of intermediate $\langle k', v' \rangle$ pairs. As soon as all chunk pairs have been processed, the map task starts a {\em local shuffle phase}, where intermediate pairs are sorted and partitioned among reduce tasks using a key hash partitioner. The same intermediate key $k'$ can be emitted by different map tasks, but ends up to be always assigned to the same reduce task. Hence, the partitioner is crucial to distribute the shuffle data to different reduce tasks so as to avoid data unbalancing issues. 

When all the local shuffle phases have terminated, MapReduce spawns  reduce tasks in parallel (we avoid discussing slow start, where reduce tasks can start fetching data earlier than this time). The number of reduce tasks is a user-defined parameter: depending on cluster capacity (and similarly to map tasks), reduce tasks can be executed in a {\em single wave} (i.e., one task per worker, all tasks in parallel) or in {\em multiple waves} (i.e., more tasks sequentialized on the same worker). The swimlanes plots in Figure~\ref{fi:ni-bs-you} show a multiple wave execution on the left and a single wave on the right.

Each reduce task starts with a local shuffle phase where $\langle k',v' \rangle$ pairs produced by different map tasks are fetched and sorted based on $k'$. Several {\em key groups} $\langle k', V_{k'}\rangle$ are obtained by shuffling, where set $V_{k'}$ contains all intermediate values associated to key $k'$. A reduce function is then invoked for each {key group} $\langle k', V_{k'}\rangle$, producing output $\langle key,value \rangle$ pairs eventually stored in the distributed file system.


\begin{figure}[t]
\centering
\includegraphics[width=0.9\columnwidth]{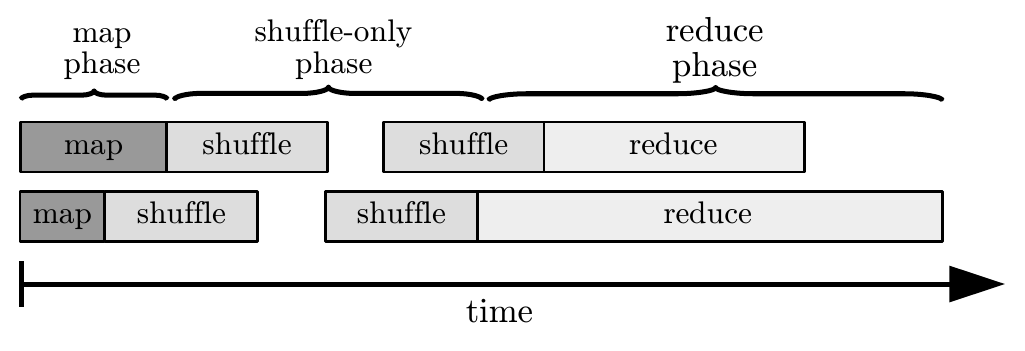}
\caption{Map, shuffle, and reduce phases in a single wave scenario with two map tasks and two reduce tasks.}
\label{fi:phases-single}
\end{figure}

\medskip
\noindent {\bf Phases.} At a higher level, the execution of a MapReduce job is commonly split into three distinct phases, roughly corresponding to the executions of map functions, shuffling, and reduce functions. As discussed before, shuffling is performed by both map and reduce tasks. Throughout this paper we assume that the {\em map phase} includes the time elapsed from the beginning of the first map function to the termination of the last one. Similarly,  the {\em reduce phase} includes the time elapsed from the beginning of the first reduce function to the termination of the last one.
The time elapsed from the termination of last map function and the beginning of the first reduce function is part of the {\em shuffle phase}. We refer to Figure~\ref{fi:phases-single} for an example. According to our definition,  during the shuffle phase the job   performs exclusively shuffling operations (i.e., no map or reduce function executions).

\medskip
\noindent {\bf Sources of skewness.}
As discussed in previous works~\cite{Gufler2011,skewtune}, several kinds of skewness can negatively impact the performance of MapReduce jobs. A first challenge is to split intermediate keys uniformly among the reduce tasks: if the hash function is not designed properly, it can introduce {\em partitioning skewness}, responsible of straggling tasks that can significantly impact the job performance. Even if keys are evenly distributed, we can have {\em shuffle data skewness} where a few key groups $(k, V_{k})$ are much larger than the other ones. This  happens quite often in real data sets and is therefore a very critical issue: e.g., nodes in large social networks exhibit power-law degree distributions, and grouping node neighbors during the map phase may result in very unbalanced neighborhood sizes~\cite{SV11}.  {\em High running times} of the reduce functions make shuffle data skewness even more critical: if the implementation of the reduce function uses a super-linear algorithm, a few executions can easily become a performance bottleneck for the entire job. 
{\em Multi-modal input distribution} can also arise when different data sets are concatenated to obtain the input of a single job: it may be the case that chunks obtained from different input data sets may require different processing times in practice due to different characteristics. We designed our model so as to take care of all these sources of skewness.  


  
\noindent 

\vspace{-2mm}
\section{The NearestFit progress indicator: theoretical model}
\label{se:methodology}

In this section we describe the overall design of our progress indicator, called {\tt NearestFit}. Similarly to previous works and according to Section~\ref{se:mapreduce}, we break down the progress of a MapReduce job into three different phases (map, shuffle, and reduce), providing separate progress estimates for each of them. 
In the description we first focus on the reduce phase: this is typically the most computationally demanding phase -- where data skewness and load unbalancing can amplify ``curse of the last reducer'' phenomena~\cite{SV11} -- and also the most complex with respect to progress prediction. The size and characteristics of its input depend indeed on the selectivity of the map function, and thus on the output of the previous phases. For the sake of presentation, we initially consider the simplified scenario where there is a single wave of reduce tasks, i.e., {\em all} reduce tasks are immediately started after shuffling. In Section~\ref{ss:model-generalization} we generalize our model by removing the single-wave assumption and by taking into account the map and shuffle phases.

\subsection{Model overview}
\label{ss:overview}

\begin{figure}[t]
\centering
\includegraphics[width=0.99\columnwidth]{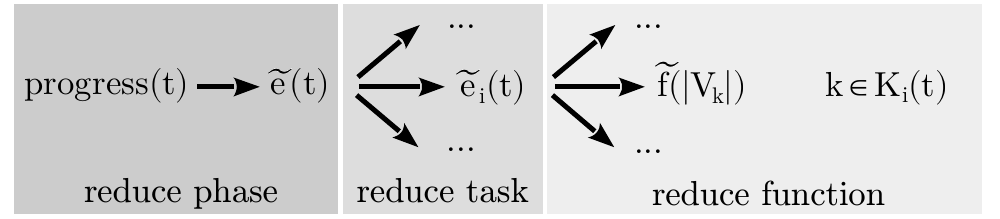}
\caption{A bird's-eye view on {\tt NearestFit}.}
\label{fig:overview}
\end{figure}

A bird's-eye view on our approach is shown in Figure~\ref{fig:overview}. For the reader's convenience, a summary of the notation introduced throughout this section  is provided in  Table~\ref{tab:notation}. When needed, we use symbol ``$\sim$'' to distinguish between the estimate of a quantity computed by our algorithm and its exact value, typically unavailable at runtime. 

\begin{table*}[t]
\begin{center}
\begin{tabular}{ |l|c| }
\hline
  
  $t$ & current time (progress is estimated at time $t$) \\\hline
  \multirow{2}{*}{$progress(t)$}  &  \multirow{2}{*}{$\left(t - t_{start}\right)\,/\,\left(\widetilde{e}(t)-t_{start}\right)$    }\\
   &   \\\hline

  \multirow{2}{*}{$\widetilde{e}(t)$}  &  \multirow{2}{*}{$ \max\limits_{reduce~tasks~i} \widetilde{e}_i(t)$} \\
   &   \\\hline

  \multirow{2}{*}{$\widetilde{e}_i(t)$}  &  \multirow{2}{*}{$p_i(t)+\widetilde{r}_i(t)$ $=$ end time estimated for reduce task $i$} \\
   &    \\\hline

  $p_i(t)$ & time of the most recent profile update for task $i$:\,\, $p_i(t)\leq t$ \\\hline
  \multirow{2}{*}{$\widetilde{r}_i(t)$} & \multirow{2}{*}{estimate of the remaining time for reduce task $i$} \\
   &   \\\hline
  $K_i$ & set of keys assigned to reduce task $i$  \\\hline
  \multirow{3}{*}{$K_i(t)$} & keys assigned to reduce task $i$ not yet processed at time $p_i(t)$  \\
   & $K_i(t)\subseteq K_i$ \\
   &  $K_i(t)=K_i$ if no profile has been collected until time $p_i(t)$ \\\hline
   $V_k$ & set of values associated to a key $k$  \\\hline
   $f(k,V_k)$ &  theoretical cost model for the reduce running time \\\hline  
  \multirow{2}{*}{${r}_i(t)$} & \multirow{2}{*}{$\sum\limits_{k\in K_i(t)}{f(k,V_k)}$= exact remaining time for reduce task $i$ at time $p_i(t)$} \\
   &    \\\hline
   $|V_k|$ &  size (in bytes) of the key group $(k,V_k)$ \\  \hline
   \multirow{2}{*}{$\widetilde{f}(|V_k|)$} &  \multirow{2}{*}{estimate of $f(k,V_k)$, as a function of the size of the input key group } \\  
    &   \\\hline  
\multirow{2}{*}{$\widetilde{r}_i(t)$} & \multirow{2}{*}{$\sum\limits_{k\in K_i(t)}{\widetilde{f}(|V_k|)}$}  \\
   &    \\\hline
\end{tabular}
\end{center}
\vspace{-2mm}
\nocaptionrule\caption{Summary of the notation used in the description of our model.}
\label{tab:notation}
\end{table*}

The  progress of the reduce phase can be computed at any point in time during its execution. In practice, updates take place at discrete moments (every 60 seconds in our implementation), rather than continuously. We denote by $progress(t)$  the progress estimated at time $t$, i.e., the percentage of elapsed time since the beginning of the phase. This is necessarily an estimate, as the wall-clock time of the reduce phase (which is required to compute the percentage) will be available only upon termination of the last reduce function, at which point the progress should be $100\%$.  

As shown in Figure~\ref{fig:overview}, we compute $progress(t)$ by estimating the ending time $\widetilde{e}(t)$ of the reduce phase. To this aim, we gather profile information at the task level, so as to predict the ending time $\widetilde{e}_i(t)$ for each reduce task $i$.
In turn, we compute $\widetilde{e}_i(t)$ by profiling the execution of the reduce functions run by task $i$ and by combining the predicted running times $\widetilde{f}(|V_k|)$. 

We now describe each step in more detail. We will discuss in Section~\ref{se:operational} profile gathering issues, describing when and which profile data are collected by {\tt NearestFit} to estimate progress.


\subsection{Estimating the reduce phase progress: $progress(t)$}
\label{ss:progress}

\begin{figure}[t]
\centering
\includegraphics[width=0.9\columnwidth]{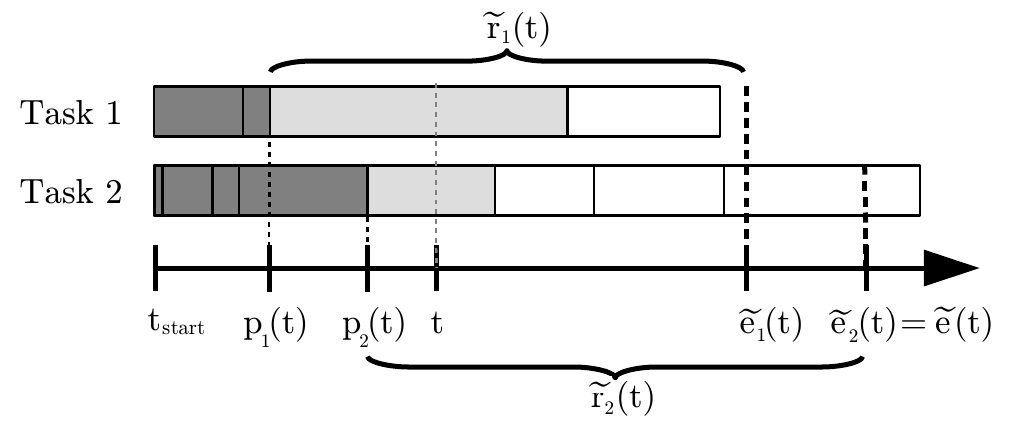}
\vspace{-1mm}
\caption{Estimated end times for two reduce tasks and for the entire reduce phase.}
\label{fi:end-time}
\end{figure}

The progress of the reduce phase is estimated at time $t$ as the ratio between the  time elapsed from the beginning of the phase until time $t$ and the predicted duration of the phase:
\begin{equation}
\label{eq:progress}
progress(t) = \frac{t - t_{start}}{ \widetilde{e}(t)-t_{start} } \times 100
\end{equation}
where $t_{start}$ is the starting time of the reduce phase (i.e., the starting time of the first reduce function) and $\widetilde{e}(t)$ is an estimate of its ending time. Since more and more profile data become available as the computation proceeds, the end time estimate is a function of $t$ and will likely return different values when computed at different times. Under the single-wave assumption, $\widetilde{e}(t)$ can be obtained as the maximum of the predicted end times among all the reduce tasks, as also shown in Figure~\ref{fi:end-time}:
\begin{equation}
\label{eq:end-time-phase}
\widetilde{e}(t) = \max\limits_{reduce~tasks~i} \widetilde{e}_i(t)
\end{equation}
The task end times $\widetilde{e}_i(t)$ are also estimates, and can be either smaller or larger than the actual end times ${e}_i(t)$. This can yield to underestimating or overestimating $\widetilde{e}(t)$. In the former case, where $\widetilde{e}(t)<e(t)$ as in the example of Figure~\ref{fi:end-time}, the estimated progress will be larger than the actual percentage of elapsed time, giving developers the temporary illusion that the computation proceeds faster than it is actually doing. The latter case is symmetric.

\subsection{Predicting the ending time  $\widetilde{e}_i(t)$ for task $i$}
\label{ss:ending-time}

The end time $\widetilde{e}_i(t)$ of a reduce task $i$ can be estimated as the sum of the {\em past}  and of the {\em remaining execution time} of task $i$. We denote the latter amount with $\widetilde{r}_i(t)$. The computation of $\widetilde{r}_i(t)$, which is addressed in Section~\ref{ss:remaining-time}, exploits profile data collected from task $i$ as described below. 


{\tt NearestFit} collects a  {\em reduce profile} for each reduce task. Profiles are periodically sent to the application master to update the progress estimate.  The reduce profile of task $i$ is updated upon termination of each reduce function invocation in task $i$, i.e., as soon as a new key group has been fully processed. 

Profile updates are unlikely to happen at the same time $t$ at which we perform a prediction: in general, the prediction at time $t$ will use profile data collected from task $i$ at some previous time, which we call $p_i(t)$. 
More formally, $p_i(t)$ is the time of the {\em most recent profile update} for task $i$ and satisfies the following properties: 
\begin{itemize}
\item by definition, $p_i(t) \leq t$;
\item due to data skewness, a reduce task may be stalled executing the same reduce instance for a long time, in which case $p_i(t)$ might be much smaller than $t$;
\item since different reduce tasks are executed in parallel without any kind of synchronization, reduce functions are likely to terminate at different times in different tasks: hence, it can be $p_i(t) \neq p_j(t)$ if $i \neq j$. 
\end{itemize}
The relation between quantities $t$, $p_i(t)$,  $\widetilde{e}_i(t)$, and $\widetilde{r}_i(t)$ for two parallel tasks is shown in Figure~\ref{fi:end-time}.
Under the single-wave assumption, we can now estimate $\widetilde{e}_i(t)$ as:
\begin{equation}
\label{eq:end-time-task}
\widetilde{e}_i(t) = p_i(t)+\widetilde{r}_i(t)
\end{equation}
Distinguishing between the prediction time $t$ and the last profile update time $p_i(t)$ implies that we can avoid speculating about the status of the reduce instance currently running at time $t$ in task $i$, which might prove to be difficult. As a consequence, in Equation~\ref{eq:end-time-task} we scale down from $t$ to $p_i(t)$ the time passed since the beginning of the task, and absorbe the time in-between $p_i(t)$ and $t$ in $\widetilde{r}_i(t)$. Figure~\ref{fi:end-time} gives a visual clue on the involved quantities.


\subsection{Estimating the remaining time $\widetilde{r}_i(t)$ for task $i$} 
\label{ss:remaining-time}

Let $K_i$ be the set of keys assigned to a reduce task $i$ (as we will see in Section~\ref{se:operational}, $K_i$ can be computed by profiling the map phase). At time $t$, $K_i$ can be conceptually partitioned into three subsets: fully processed keys (whose corresponding reduce instances have already terminated), untouched keys (whose corresponding reduce instances have yet to be started), and a single key that is currently being processed (whose corresponding reduce instance started at time $p_i(t)$ and has not yet terminated). We denote with $K_i(t)\subseteq K_i$ the set of currently processed and untouched keys at time $t$ (light gray and white rectangles in Figure~\ref{fi:end-time}). 
%
%
The exact remaining running time $r_i(t)$ for task $i$ is now given by:
\begin{equation}
\label{eq:exact-remaining-time}
r_i(t)=\sum\limits_{k\in K_i(t)}{f(k,V_k)}
\vspace{-1mm}
\end{equation}
where $f(k,V_k)$ is a {\em cost model} for the reduce running time, depending both on the input key $k$ and on the input values $V_k$. Since the true cost model $f$ is typically unknown, we need to estimate $f(k,V_k)$. Our approach is to {\em learn from past executions} of the reduce functions: since we know exactly $f(k,V_k)$ for each fully processed key $k\in K_i\setminus K_i(t)$, we  exploit this knowledge to predict the running time of the unprocessed keys in $K_i(t)$. 

With a slight abuse of notation, we will use $|V_k|$ to denote the size (in bytes) of the key group $(k,V_k)$. Guided by asymptotic analysis and by previous works on performance profiling (see, e.g.,~\cite{CDF12,GAW07}), we assume that the running time of a reduce function depends on the input size, and not on the actual input values: thus $f(k_1,V_{k_1})\approx f(k_2,V_{k_2})$ whenever $|V_{k_1}|\approx |V_{k_2}|$.
With this assumption, an estimate $\widetilde{r}_i(t)$ for the remaining running time of reduce task $i$ can be obtained as:
\begin{equation}
\label{eq:approx-remaining-time}
\widetilde{r}_i(t) = \sum\limits_{k\in K_i(t)}{\widetilde{f}(|V_k|)}
\vspace{-1mm}
\end{equation}
where $\widetilde{f}(|V_k|)$ is an estimate of $f(k,V_k)$, depending on the input size. In Section~\ref{ss:nearestfit} we discuss how to compute $\widetilde{f}(|V_k|)$. 


\subsection{Predicting the running time $\widetilde{f}(|V_k|)$ of reduce functions} 
\label{ss:nearestfit}

We use regression analysis to predict the running time of a reduce function on key group $\langle k,V_k\rangle$, using $|V_k|$ as independent variable. In particular, we combine two well-known techniques: {\em $\delta$-nearest neighbor regression} and {\em curve fitting}. Each technique suits a different scenario: we will see in Section~\ref{se:experiments} that a careful integration yields very accurate progress estimates, even in highly skewed settings.

\begin{figure}[t]
\centering
\includegraphics[width=0.7\columnwidth]{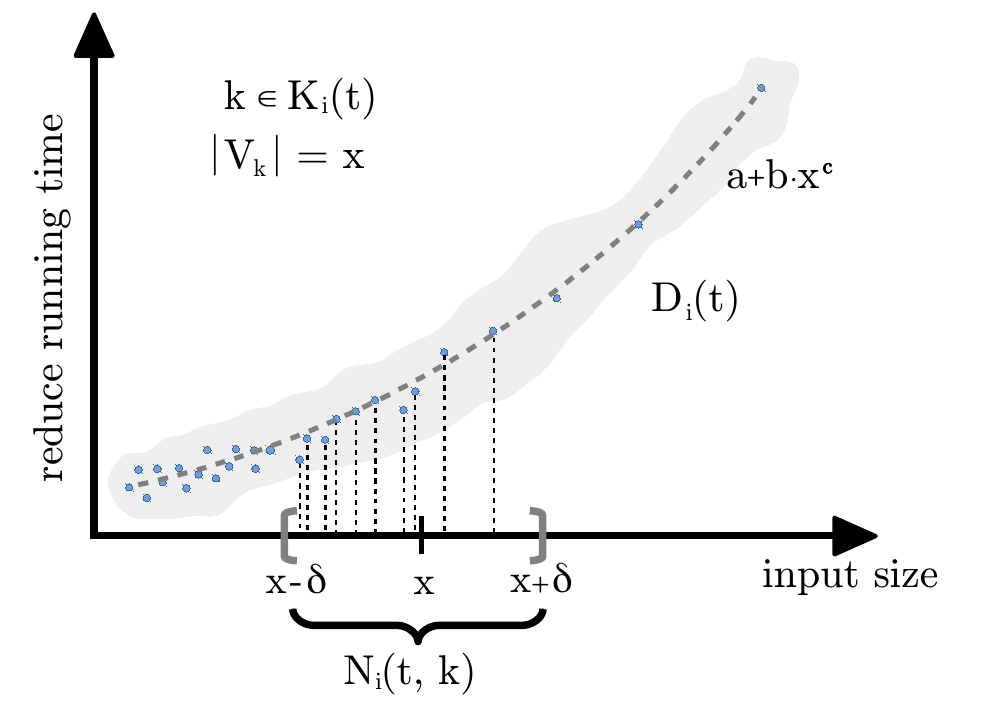}
\caption{Profile data used in the prediction of $\widetilde{f}(|V_k|)$: $D_i(t)$, $\delta$-neighborhood  $N_i(k,t)$, and a curve fitting model.}
\label{fi:datapoints}
\end{figure}

\paragraph{Nearest-neighbor regression.} This is a kind of instance-based learning, where data is classified according to the training  examples that are closest to the new point~\cite{LRU14}. In our setting, the training examples are the fully processed keys $k' \in K_i\setminus K_i(t)$, for which the exact running time $f(k',V_{k'})$ is known. 

Given an unprocessed key $k\in K_i(t)$, we can predict $f(k,V_{k})$ as follows. Let $\delta\ge 0$ be a  constant and let $N_i(t, k)\subseteq K_i\setminus K_i(t)$ be the $\delta$-neighbourhood of $k$: $N_i(t, k)$ contains those keys $k'$ that are fully processed at  time $t$ and whose input size is $\delta$-close to $k$ (formally, $|V_{k'}| \in [|V_k| - \delta, |V_k| + \delta]$). Different rules can be used to derive $\widetilde{f}(|V_k|)$ from $N_i(t, k)$. The approach we take is among the simplest ones and is to average the running times observed for the $\delta$-neighborhood:
\vspace{-2mm}
\begin{equation}
\label{eq:nearest-neighbor}
\widetilde{f}(|V_k|) = \frac{\sum\limits_{k'\in N_i(t, k)}{f(k',V_{k'})}}{|N_i(t, k)|}
\end{equation}
Nearest-neighbor regression is well-defined and yields an estimate for $\widetilde{f}(|V_k|)$ only if $N_i(t, k)$ is non-empty. In some cases, however, it may be necessary to perform significant {\em extrapolations}  during the reduce phase, i.e., to predict the running times well-beyond the range of the already observed input sizes. This is typically the case in the presence of data skewness, where a few reduce instances get inputs much larger than the average input size: these straggling instances prominently affect the task ending time, but are likely to have an empty neighborhood.

\paragraph{Curve fitting.} Statistical curve fitting addresses the construction of a mathematical model that has the best fit to a set of data points~\cite{CCKT83}. In our setting, the set $D_i(t)$ of data points used to build the model at time $t$ has two dimensions, describing respectively input sizes and running times of terminated executions of the reduce function:
\begin{equation}
\label{eq:data-points}
D_i(t)=\bigcup\limits_{k'\in K_i\setminus K_i(t)} \{\,(|V_{k'}|, f(k',V_{k'}))\,\}
\end{equation}
Figure~\ref{fi:datapoints} shows the relation between $D_i(t)$ and the $\delta$-neighborhood $N_i(t, k)$ of an unprocessed key $k$. As a fitting model, we use $a + b \cdot x^c$, which generalizes both power law and linear models (where $a=0$ and $c=1$, respectively). As shown in previous works~\cite{CDF12,GAW07}, this model can characterize different kinds of real-world benchmarks. Curve fitting yields a closed form expression that approximately describes the running time of a reduce function on input size $|V_k|$, yielding clues to its growth rate. It will predict $\widetilde{f}(|V_k|)$ as:
\vspace{-1mm}
\begin{equation}
\label{eq:curve-fitting}
\widetilde{f}(|V_k|) = a + b \cdot |V_k|^c
\end{equation}
Curve fitting is quite appealing, but can be difficult to tune in practice due to data noise, often producing low-quality cost models (with small $R^2$ values). Moreover, it can be unstable when repeatedly used over time, as in progress analysis: even small changes in the model coefficients $a$, $b$, and $c$ can unfortunately yield quite different predictions.

\paragraph{NearestFit: combining nearest neighbors with curve fitting.} We observed that nearest-neighbor regression is typically very accurate, but cannot  be always applied. Conversely, curve fitting is potentially always applicable, but its accuracy crucially depends on the fitting quality. We therefore combined the two techniques so as to overcome their drawbacks while retaining their advantages. To predict $\widetilde{f}(|V_k|)$, the reduce task $i$ orderly considers the following options:
\begin{enumerate}
    \item checks if $N_i(t, k)$ is non-empty: in this case uses the nearest-neighbor prediction according to Equation~\ref{eq:nearest-neighbor};
    \item computes a model $a + b \cdot x^c$ on set $D_i(t)$ specified in Equation~\ref{eq:data-points}:  if the model has good fitting quality ($R^2 \geq 0.9$), uses the prediction given by the model;
    \item uses nearest-neighbor regression, if possible, on data points collected from reduce executions in {\em all} reduce tasks;
    \item among the curve fitting models available for {\em all} reduce tasks, predicts according to the model that best estimates the data points $D_i(t)$ currently available at task $i$.
\end{enumerate}
Notice that the first two steps use task-local information, while the latter ones exploit information gathered from all the other tasks. Several other combination strategies could be considered. In our experiments, this approach provided very  reasonable tradeoffs between accuracy and efficiency. Indeed, it resorts to curve fitting -- and ultimately to global information gathered from other tasks -- only in very rare cases. Dealing with these cases is however crucial for an accurate progress estimate, since they often correspond to the most time-demanding (and difficult to predict) executions. 


\subsection{Generalizations}
\label{ss:model-generalization}

\paragraph{Map phase.} The approach described for the reduce phase can be also extended to the map phase. From a high level perspective, {\tt NearestFit} exploits two main ingredients: the running times of past function executions and a characterization of the input. Running times can be collected for map functions similarly to the reduce phase. On the other hand, the main challenge is how to compute the set $K_j$ of $\langle key,value\rangle$ pairs assigned to a map task $j$: while $K_i$, for each  reduce task $i$, can be obtained by profiling the map phase (as we will see in Section~\ref{se:operational}), we have no such information about map input before actually running the map tasks. Scanning the entire job input before the execution would be prohibitive. Hence, similarly to previous works~\cite{SMHMYL13,Le2014} we can exploit sampling to approximate this information. 

In practice, stragglers are  unlikely in the map phase: splits assigned to a map task have a limited size, and the input of each map function is simply a single $\langle key,value\rangle$ pair, instead of the possibly long list of values received by reduce functions. Hence, in practice a linear prediction model is good enough to estimate progress of the map phase. This is the choice in our implementation, which proved to be fully satisfactory in our experiments.

As a final note, the output of map functions could be locally aggregated through the use of combiners, which however operate on a limited amount of data: the output of map functions is buffered, and buffers typically have a small size. Moreover, combiners run in parallel with the map functions, as a separate thread during the map phase. Hence, we do not explicitly predict the running time of combiners, which is covered by the map phase progress.

\paragraph{Shuffle phase.} As observed in previous works~\cite{microbench}, the shuffling phase is application-independent and its performance can be safely estimated using only execution profiles collected on micro-benchmarks. Microbenchmark predictions need to be scaled proportionally to the amount of shuffle data, which can be obtained by profiling map tasks.

\paragraph{Removing the single-wave assumption.}
So far, we have assumed that all reduce tasks immediately start after shuffling, i.e., at time $t_{start}$. In general, this is not true if the task parallelism available in the cluster is smaller than the number of reduce tasks. In order to consider multiple execution waves in our prediction model, we need to refine Equation~\ref{eq:end-time-task}, which estimates the ending time $\widetilde{e}_i(t)$ of a reduce task $i$. At any time $t$, we distinguish between unscheduled and running tasks (the ending time for terminated tasks is already known).

For running tasks, everything is estimated as in Section~\ref{ss:ending-time}. However, before receiving the first profile from task $i$, we take care of initializing $p_i(t)$ to the task starting time (instead of $t_{start}$).

For unscheduled tasks, we first estimate their remaining time $\widetilde{r}_i(t)$, including the running time required for shuffling pairs (this is done similarly to the shuffle phase). Since local task information is not yet available, we exploit information collected from the other tasks, predicting running times according to points 3 and 4 of the combined approach described in Section~\ref{ss:nearestfit} (in point 4, we choose the curve fitting model with the best quality $R^2$). Once $\widetilde{r}_i(t)$ is available for all tasks, we simulate the scheduler to deduce the starting time for each unscheduled task. We thus obtain an estimate of $p_i(t)$. We remark that reasoning on the scheduler choices  is a standard practice in previous performance analysis papers (see, e.g.,~\cite{HB13,ARIA11,VCC11,parallax}), which typically assume a simple online greedy scheduling strategy as in this paper.

\section{An operational view of NearestFit}
\label{se:operational}

To estimate progress, {\tt NearestFit} needs to collect diverse information, combining coarse- and fine-grained profiles. In this section we discuss the orchestration of {\tt NearestFit} profiling components, which work in different job phases and exploit both distributed and centralized computations. We remark that gathering fine-grained profile data can result in very large time and space overheads,  being thus unfeasible in practice. For the time being we ignore this issue, describing a simple-minded operational view of the theoretical model presented in Section~\ref{se:methodology}. We defer to Section~\ref{se:practical} the discussion of algorithmic techniques aimed at making the model practical: this will be achieved by computing suitable approximations for some of the quantities explicitly maintained throughout this section. 

\subsection{Gathering profiles}
\label{ss:gathering-naif}

{\tt NearestFit} uses three different profiles: (1) {\em map task profiles} collected during the map phase by each map task; (2) a {\em key distribution profile} computed by the application master from the map task profiles; and (3) {\em reduce task profiles} collected during the reduce phase by each reduce task.

\smallskip
\noindent {\bf\em Map task profiles.} For each key $k$ emitted by a map task $j$, $k$ may be produced, associated with different values, by  distinct map function invocations within task $j$. Let $V_{k,j}$ be the set of values associated to $k$ by task $j$. The map task profile maintains all pairs $(k,|V_{k,j}|)$. The profile is sent to the application master upon termination of the map task.

\smallskip
\noindent {\bf\em Key distribution profile.} For each key $k$ emitted during the map phase, the application master is aware of which reduce task will be responsible of processing $k$. By reversing this information, it can therefore obtain the set $K_i$ of keys assigned to each reduce task $i$. For each $k\in K_i$, we also need to know the size $|V_{k}|$ of its key group in order to compute $\widetilde{r}_i(t)$ as in Equation~\ref{eq:approx-remaining-time}. $|V_{k}|$, however, is unknown before fetching and merging the sets $V_{k,j}$ stored in the local file systems of each map worker $j$. The map task profiles are thus put to use to compute $|V_{k}|$:
\begin{equation}
\label{eq:size}
|V_{k}|=\sum_{map~tasks~j}|V_{k,j}|
\end{equation}
Overall, the key distribution profile will contain, for each reduce task $i$, all the keys $k\in K_i$ along with their value sizes $|V_k|$.

\smallskip
\noindent {\bf\em Reduce task profiles.} Reduce task profiles are collected as described in Section~\ref{ss:ending-time}. Profile data gathered at time $p_i(t)$ contain information on the reduce function execution just completed in reduce task $i$. If $\langle k,V_k\rangle$ is the key group processed by that reduce function, the profile reports the key $k$, the input size $|V_k|$, and the running time $f(k,V_k)$. 

\subsection{Application master data structures}
\label{ss:data-structures-naif}

During the reduce phase,  for each reduce task $i$ the application master maintains: 
\begin{itemize}
\item the last reduce profile update time $p_i(t)$ -- see Section~\ref{ss:ending-time}; 
\item the set of unprocessed keys $K_i(t)$ -- see Section~\ref{ss:remaining-time}; 
\item the past executions data points $D_i(t)$ -- see Section~\ref{ss:nearestfit}. 
\end{itemize}
At time $t_{start}$, we have $K_i(t)=K_i$, $p_i(t)=t_{start}$, and $D_i(t)=\emptyset$. $K_i$ is obtained from the key distribution profile. When the reduce function for key group $\langle k,V_k\rangle$ terminates at task $i$, the application master updates $p_i(t)$ to the termination time, removes $k$ from $K_i(t)$, and adds a new point $(|V_k|,f(k,V_k))$ to $D_i(t)$. 

\subsection{Updating progress estimates}
\label{ss:update-progress-naif}

At any time $t$, the progress indicator can be brought up to date by first computing $\widetilde{f}(|V_k|)$ for each unprocessed key $k$ as follows. The curve fitting model parameters of each reduce task are  updated using sets $D_i(t)$. Moreover, the $\delta$-neighborhood $N_i(t,k)$ of keys $k\in K_i(t)$ is obtained from $|V_k|$ and $D_i(t)$ (see Figure~\ref{fi:datapoints}) and Equation~\ref{eq:nearest-neighbor} is used if the $\delta$-neighborhood is non-empty. Once estimates $\widetilde{f}(|V_k|)$ are available, the progress can be determined according to Equations~\ref{eq:approx-remaining-time},~\ref{eq:end-time-task}, \ref{eq:end-time-phase}, and \ref{eq:progress}.

\section{Making NearestFit practical}
\label{se:practical}

As observed in Section~\ref{se:operational}, we need to minimize the amount of profile data that percolates through the framework -- and ultimately through the network -- both for time and for space efficiency reasons. In practice, we cannot afford to  collect on worker nodes fine-grained profiles that are later processed by the application master in a centralized way. The key insight to solve this issue is to get rid of keys in our profiles. As shown by Equation~\ref{eq:approx-remaining-time}, only input sizes $|V_k|$ are needed to predict the remaining times: keys are not used except for defining the terms of the sum, which iterates over $k\in K_i(t)$. In this section we show that we can avoid to maintain $K_i(t)$ explicitly, revisiting the operational view of {\tt NearestFit} presented in Section~\ref{se:operational}.
Getting rid of keys presents challenges at different points, but also offers several benefits: the set of distinct keys in a MapReduce job is typically huge, while the set of distinct input sizes is much smaller and tractable (many key groups have similar, if not identical sizes, and can be aggregated). 

\subsection{Gathering key-independent profiles}
\label{ss:gathering-opt}

\noindent {\bf\em Map task profiles.} A map task $j$ splits its emitted keys into two logical categories: {\em explicit} and {\em implicit} keys. We recall that $V_{k,j}$ is the set of values associated to key $k$ by task $j$. Pairs $(k,|V_{k,j}|)$ are maintained in the profile only for explicit keys, whose number is fixed to a constant $\lambda$ (in our implementation $\lambda = 2000$). Explicit keys are the $\lambda$ keys with the largest sizes $|V_{k, j}|$. This set can be obtained at the end of the local shuffle phase of a map task. Intuitively, we focus on explicit keys because large numbers of values can result in high reduce running times. Slow reduce instances, in turn, are likely to have non-negligible effects on the job progress.

The remaining (implicit) keys are not reported in the map task profile, except for aggregate values. Let $K_{i,j}$ be the set of keys emitted by map task $j$ and assigned to reduce task $i$, and let $E_{i,j}$ be the maximal subset of $K_{i,j}$ containing only explicit keys. Then, for each reduce task $i$, the map task profile contains the pair:
\begin{equation}
\label{eq:implicit-map-pairs}
(|K_{i,j}| - |E_{i,j}|, \sum\limits_{k \in K_{i,j} \setminus E_{i,j}} {|V_{k, j}|})
\vspace{-1mm}
\end{equation}
This gives the number of implicit keys emitted by map task $j$ and assigned to reduce task $i$ and their total size.

\medskip
\noindent {\bf\em Key distribution profile.} Since we no longer have information on all keys, we cannot compute $K_i$ explicitly. We operate differently on explicit and implicit keys. 

For implicit keys assigned to reduce task $i$, we can compute their total size by summing up the values $\sum{|V_{k, j}|}$ received by all map tasks $j$. Instead, we cannot know their number, since summing up the numbers $|K_{i,j}| - |E_{i,j}|$ received by map tasks $j$ yields a very inaccurate upper bound: some of the map tasks might emit the same key, that would be counted more than once in the sum. We will later show how to get a safe estimate using information collected during the reduce phase (see Section~\ref{ss:update-progress-opt}).

We now consider explicit keys received by reduce task $i$. For each $k\in E_{i,j}$, we know from the map task profiles its number of values $|V_{k, j}|$ emitted by map task $j$. We thus estimate:
\begin{equation}
\label{eq:approx-explicit-size}
\widetilde{|V_k|} = \sum\limits_{map~task~j~\,}{\sum\limits_{k \in E_{i,j}} |V_{k, j}|}
\vspace{-1.5mm}
\end{equation}
This is not the exact value size of key $k$: $\widetilde{|V_k|} \leq |V_k|$ since the same key can be explicit in a map task $j$ and implicit in a different map task. Map tasks for which $k$ is implicit do not contribute to the sum. However, in the presence of data skewness, we expect $|V_{k}|$ to be large with respect to the majority of value sizes of other keys, and therefore it is likely that $\widetilde{|V_k|} \simeq |V_k|$.

To bound memory consumption of the application master, we introduce an additional approximation when computing the sum in Equation~\ref{eq:approx-explicit-size}. 
The master must merge the explicit keys emitted by all map tasks in order to estimate ${|V_k|}$ according to Equation~\ref{eq:approx-explicit-size}. 
Although the number $\lambda$ of explicit keys emitted by each map task is bounded, a large number of map tasks might yield a prohibitively large set of explicit keys on the master side. For instance, if there are $50000$ map tasks and each of them emits $2000$ explicit keys, the master would receive one million pairs. To overcome this issue, we aggregate the sizes of the smallest explicit keys in the union set, making them implicit. This is done through a space-efficient algorithm that maintains frequent items over data streams~\cite{spaceSaving}, where the sets of explicit keys received from each map task are regarded as streams. The Space Saving algorithm~\cite{spaceSaving} processes  streams on-the-fly, without storing them entirely, and can return the heaviest keys -- as well as an estimate of their sizes -- with very high accuracy.

\vspace{-0.5mm}

\medskip
\noindent {\bf\em Reduce task profiles.} Reduce task profiles are computed as described in Section~\ref{ss:gathering-naif}, but omitting the input key: for each key group $\langle k,V_k\rangle$ processed by a reduce function, the profile reports only $|V_k|$ and the running time required to process the group.

\subsection{Application master data structures}
\label{ss:data-structures-opt}

With respect to the data structures used in Section~\ref{ss:data-structures-naif}, we can no longer maintain the set $K_i(t)$ of  keys not yet processed by reduce task $i$ at time $t$. Indeed, $K_i(t)$ was initialised to $K_i$, which is now unknown to the application master except for explicit keys.
Instead of $K_i(t)$, at time $t$ we now maintain:
\begin{itemize}
\item a set $S_i(t)$ of approximate sizes of unprocessed explicit keys;
\item the total size $s(t)$ of unprocessed implicit keys.
\end{itemize}
Both $S_i(t)$ and $s(t)$ can be initialised with information available in the map distribution profile (recall that the approximate sizes of explicit keys have been computed in Equation~\ref{eq:approx-explicit-size}).

When the reduce function for key group $\langle k,V_k\rangle$ terminates at reduce task $i$, the application master receives $(|V_k|,f(k,V_k))$ in the reduce task profiles. $S_i(t)$ and $s(t)$ are updated as follows:
\begin{itemize}
\item if $|V_k|$ is (close to) an explicit size $\widetilde{|V_k|}$ maintained in $S_i(t)$, we delete $\widetilde{|V_k|}$ from $S_i(t)$;
\item otherwise we decrease $s(t)$ by $|V_k|$.
\end{itemize}
%
%
$D_i(t)$ and $p_i(t)$ can be updated as before (see Section~\ref{ss:data-structures-naif}). We remark that all the data structures of the application master are now independent of keys, and use only the sizes of the key groups.

\subsection{Updating progress estimates}
\label{ss:update-progress-opt}

We now discuss how to update the progress indicator using the data structures described in Section~\ref{ss:data-structures-opt}. The main issue is to compute the remaining time of each reduce task $i$: Equation~\ref{eq:approx-remaining-time} iterates over the set $K_i(t)$ of unprocessed keys, which is however no longer available.
This issue can be naturally solved for explicit keys, whose approximate sizes are explicitly maintained in $S_i(t)$. The {\em remaining time for explicit keys} can be computed as:
\begin{equation}
\label{eq:remaining-time-explicit}
\widetilde{r}_i(t) = \sum\limits_{\widetilde{|V_k|}\in S_i(t)}{\widetilde{f}(\widetilde{|V_k|})}
\vspace{-1mm}
\end{equation}
where $\widetilde{f}$ is obtained as before either via nearest-neighbor regression or via statistical curve fitting.

Computing the {\em remaining time for implicit keys} is more challenging, since we know only their total size $s(t)$. Computing the average  size for implicit keys turns out to be impossible, as the number of implicit keys is unknown  (see Section~\ref{ss:gathering-opt}). With no information on individual sizes $|V_k|$ and on the number of distinct implicit keys assigned to a reduce task, we cannot decide which and even how many terms contribute to the summation given by Equation~\ref{eq:approx-remaining-time}. 
We approximate the missing information by learning the key size distribution during the execution of reduce functions. Namely, we maintain an approximate histogram of the most frequent sizes encountered during the reduce phase and we partition $s(t)$ based on this distribution. In this way we obtain estimates of the number of implicit keys and of their sizes, which we plug in  Equation~\ref{eq:approx-remaining-time} to obtain the remaining time for implicit keys.

Notice that throughout the entire computation we never used keys (neither explicit nor implicit), but we limit to exploit different -- explicit or aggregate -- approximations of the key group sizes.

\section{Hadoop implementation}
\label{se:implementation}

We implemented {\tt NearestFit} on top of Hadoop 2.6.0. Apache Hadoop~\cite{hadoop} is perhaps the most widespread open-source implementation of MapReduce. The implementation consists of three main components: {\em map task tracker}, {\em reduce task tracker}, and {\em progress monitor}. The map task tracker is deployed on map worker nodes and generates a map task profile once and for all during the local shuffling phase. Similarly, the reduce task tracker generates a reduce task profile during the execution of each reduce task. Reduce profiles are updated during the computation, and updates are periodically sent to the progress monitor. However, differently from the theoretical description of Section~\ref{ss:ending-time}, to reduce communication overhead the reduce task tracker uses a buffering strategy, delaying updates until the termination of a group of reduce functions. The progress monitor fetches map and reduce profiles, updating its data structures. Notice that the progress monitor is a centralized component. For performance reasons, it is deployed as a service inside the same application master launched by Hadoop for handling the job execution. We now discuss some relevant aspects and optimizations of our implementation.

\paragraph{Measuring value size.} In our model we repeatedly used the size (in bytes) required for storing $\langle key, value\rangle$ pairs. Since keys and values have to be (de)serialized by Hadoop, our implementation can efficiently collect their size, incurring a negligible overhead.

\paragraph{Key hashing.} The key emitted by a map function in Hadoop can be any kind of user-defined Java class object. To avoid  high processing and communication costs due to large keys, our implementation considers a 64-bit hash value computed on the serialized representation of the key, instead of its actual object representation. This is more space-efficient and incurs no major drawback (e.g., hash conflicts) in practice.

\paragraph{Noise and communication reduction via smoothing.} 
Working on sizes instead of keys, as described in Section~\ref{se:practical}, makes it possible to perform {\em smoothing}, i.e., to aggregate information about similar executions, characterized by similar running times and similar input sizes. In our implementation we aggregate data points of $D_i(t)$ by merging past executions of reduce functions with the same input size and whose running times differ by at most $500$ ms. Smoothing turns out to be crucial to bound the amount of profile data sent to the progress monitor and also to reduce data noise, which could easily harm curve fitting algorithms. 


\paragraph{Measuring running times.} Measuring the running time of each reduce function execution may be inaccurate for short runs and cause a non-negligible slowdown. In our first implementation of {\tt NearestFit}, we experienced a slowdown up to $30\%$ w.r.t. native execution, most prominently on jobs characterized by many short reduce runs. We have thus introduced a {\em bursting} strategy driven by the key group sizes $|V_k|$. The main idea is to obtain accurate measurements for key groups related to (heavy) explicit keys, while avoiding frequent measurements for key groups related to (lightweight) implicit keys. In more details, our bursting strategy delivers a time measurement only if one of the following conditions is met:
\begin{itemize} 
    \item the size of the key group currently being processed is larger than an implementation-dependent threshold size;
    \item the number of reduce executions that have been {\em skipped} (i.e., not explicitly measured) is larger than another implementation-dependent threshold.
\end{itemize}
The first condition is checked during the invocation of the {\tt next} operator, i.e., while iterating over the list of values passed to a reduce function. The second condition is checked after each reduce execution. Thresholds can be customized:  in our experiments we used 50 bytes and 100 skipped executions, respectively. 

Let $t$ be the cumulative time of a burst of consecutive reduce executions. At the end of the burst we need to assign $t$ to the different executions: instead of partitioning the cumulative time uniformly, our approach is to split $t$ proportionally to the input size of each invocation. In practice, we are using a linear model here, but only for executions that  are very short. 

\section{Experimental setup}
\label{se:setup}

In this section we describe the experimental framework that we used to evaluate {\tt NearestFit}. We discuss the inner workings of state-of-the-art competitors, synthetic and real benchmarks used in our empirical assessment, performance metrics, and platform setup.

\subsection{State-of-the-art progress indicators}
\label{ss:competitors}

We compared {\tt NearestFit} against three different progress indicators: one is the standard indicator provided by Hadoop~\cite{hadoop}, while the last two exploit techniques presented in previous works~\cite{parallax,VCC11}. We implemented the latter techniques doing our best to faithfully convey in the implementation their key ideas. The description below is cast into our notation and focuses on the reduce phase.

\paragraph{Hadoop progress indicator.} Progress estimates in Hadoop do not take into account running times. The progress of the reduce phase is given by the average progress of the reduce tasks, and the progress of each task is simply the percentage of shuffle data read by the task itself. 
Using the average can badly affect progress estimates in the presence of load unbalancing.

%

\paragraph{Job ratio.} The main idea in~\cite{parallax} and~\cite{VCC11} is to compute an {\em average execution speed} $\alpha(t)$ for the reduce functions across all the reduce tasks. With the notation introduced in Section~\ref{se:methodology}:
\begin{equation}
\label{eq:job-ratio}
\alpha(t)=\frac{\sum\limits_{tasks~i~~}{\sum\limits_{k\in K_i\setminus K_i(t)}{f(k,V_k)}}}{\sum\limits_{tasks~i~~}{\sum\limits_{k\in K_i\setminus K_i(t)}{|V_k|}}}
\end{equation}
An exponentially weighted moving average (EWMA) could be applied to $\alpha(t)$ in order to smooth fast variations over time~\cite{parallax}. The remaining time $\widetilde{r}_i(t)$ for reduce task $i$ depends on $\alpha(t)$ as follows:
\begin{equation}
\label{eq:remaining-job-ratio}
\widetilde{r}_i(t)=\alpha(t)\times {\sum\limits_{k\in K_i(t)}{|V_k|}}
\end{equation}
Equations~\ref{eq:job-ratio} and \ref{eq:remaining-job-ratio} can be efficiently computed by collecting aggregate profile data at the task level. 
Job ratio correctly takes into account load unbalancing among different reduce tasks. However, by computing the average across all tasks, it implicitly assumes that different tasks process input data at a similar speed.

\paragraph{Task ratio.} This is a variant of job ratio that computes a distinct average execution speed $\alpha_i(t)$ per task, with the goal of addressing different behaviors of the reduce tasks. As long as the amount of data processed by task $i$ is small, $\alpha_i(t)=\alpha(t)$. Otherwise:
\begin{equation}
\label{eq:task-ratio}
\alpha_i(t)=\frac{{\sum\limits_{k\in K_i\setminus K_i(t)}{f(k,V_k)}}}{{\sum\limits_{k\in K_i\setminus K_i(t)}{|V_k|}}}
\end{equation}
Similarly to job ratio, an exponentially weighted moving average can be applied to values $\alpha_i(t)$ to smooth fast variations. The remaining time $\widetilde{r}_i(t)$ for reduce task $i$ can be now predicted as:
\begin{equation}
\label{eq:remaining-task-ratio}
\widetilde{r}_i(t)=\alpha_i(t)\times {\sum\limits_{k\in K_i(t)}{|V_k|}}
\end{equation}
This model takes into account both load unbalancing and different execution speeds of the reduce tasks. However, it still assumes that the running time of a reduce task scales {\em linearly} with respect to the size of its input data, ignoring any issues given by superlinear reduce computations.

\subsection{A synthetic benchmark}

To perform a preliminary empirical assessment of the accuracy of different progress indicators, we implemented a synthetic MapReduce job where we can customize the running time of the reduce function, choosing among $\Theta(n)$, $\Theta(n^2)$, $\Theta(n^3)$, or $\Theta(n^4)$ (linear, quadratic, cubic, or quartic). Data sets for the synthetic benchmark are generated using different skewness levels, where skewness is specified by a number $\sigma>1$: if $n$ is the largest input size of a reduce function, we appropriately generate keys so that approximately $\sigma^i$ functions have input size $n/\sigma^i$, for $i\ge 0$. 

\begin{table*}
\begin{footnotesize}
\begin{center}
\begin{tabular}{ |l||c||c||c| }
\hline
    {\sc benchmark} & {\sc goal} & {\sc reduce complexity} & \# {\sc datasets} \\\hline\hline
    {\tt WordCount} & word frequency counter & \multirow{2}{*}{$\Theta(n)$} & \multirow{2}{*}{1}\\
    {\tt WordCount-NC} & word frequency counter (no combine) & & \\\hline
    {\tt InvertedIndex} & computing inverted index of a set of documents & $\Theta(n^2)$ & 2\\\hline
    {\tt 2PathGenerator} & two-length path generation in a graph & $\Theta(n^2)$ & 6 \\\hline
    {\tt TriangleCount} & triangle counting in a graph & $\Theta(n)$ & 6\\\hline
    {\tt NaturalJoin} & natural join $R\bowtie S$ between relations $R$ and $S$ & $\Theta(n_R(k) \cdot n_S(k))$ & 5 \\\hline
    {\tt MatMult-Opt} & sparse matrix multiplication using an adaptive partitioner &  &  \\
    {\tt MatMult-Rnd} & sparse matrix multiplication using a random partitioner & $\Theta(n^3 \cdot d \cdot d')$ & 2 \\
    {\tt MatMult-Unbal} & sparse matrix multiplication using an unbalanced partitioner &  &  \\\hline
\end{tabular}
\end{center}
\end{footnotesize}
\vspace{-2mm}\nocaptionrule\caption{A summary of the real-world benchmarks considered in our experimental evaluation.}
\label{ta:benchmarks}
\end{table*}

\subsection{Real-world benchmarks}
\label{benchmarks}

We have carefully chosen a variety of real-world benchmarks, whose main features are summarized in Table~\ref{ta:benchmarks}, trying to expose different computational patterns typical of MapReduce applications.

\smallskip
\noindent $\bullet$~~{\tt WordCount} is probably the most well-known MapReduce example. It counts word frequencies inside a set of documents. After tokenizing each document into separate lines during the map phase, each reduce function invocation computes the frequency of a different word. Both map and reduce functions have a linear complexity.  By default combiners are active. {\tt WordCount-NC} is the variant in which combiners are disabled. As input dataset for these applications we used a 50GB archive of Wikipedia articles. 

    
\smallskip
\noindent $\bullet$~~{\tt InvertedIndex} is taken from the PumaBenchmark suite~\cite{puma12}. It computes the inverted index of a set of documents. The map function tokenizes lines of a document $d$, emitting for each word the pair (word, $d$), in linear time. The reduce function is the identity function, except for eliminating duplicates in the output. Due to a sub-optimal choice of data structures for duplicate detection, the reduce function has quadratic running time. Data sets have been derived from the 50GB Wikipedia archive, arbitrarily partitioning articles into either 5000 large documents or 50000 small documents.

\smallskip
\noindent $\bullet$~~{\tt 2PathGenerator} generates paths of length two in a graph. This is a common step in many graph analytics applications (see, e.g.,~\cite{FFF14,SV11,LRU14}). Our code is taken from~\cite{FFF14}. Map functions require constant time and emit a pair $\langle u,v\rangle$ for each arc $(u,v)$ in the graph. Reduce functions, given the neighborhood of a node $u$, emit a quadratic number of length-2 paths centered at $u$. Data sets used for this benchmark are six different social networks and web graphs taken from the Stanford Network Analysis Project~\cite{snap}.
    
\smallskip
\noindent $\bullet$~~{\tt TriangleCount} implements round 2 of the {\tt NodeIterator} algorithm described in~\cite{SV11}. The reduce function receives a pair of nodes $u$ and $v$ as key and a list of  neighbors common to $u$ and $v$ as values. If arc $(u,v)$ exists, for each common neighbor $w$ it emits a pair $\langle w,1\rangle$. This requires linear time. At the end of the round, the number of emitted pairs is six times the number of triangles in the graph (see~\cite{SV11} for details). We tested {\tt TriangleCount} on the SNAP graphs~\cite{snap}.

\smallskip
\noindent $\bullet$~~{\tt NaturalJoin} is a naive MapReduce implementation of the natural join operator between two relations $R$ and $S$~\cite{LRU14}. Both $R$ and $S$ consist of tuples with two attributes. The map function, for each tuple, emits the first attribute as key and the second attribute as value. The reduce function joins values associated to same key, computing the Cartesian product between input tuples. If $n_R(k)$ and $n_S(k)$ denote the number of tuples with key $k$ in relations $R$ and $S$, respectively, the reduce function has complexity $\Theta(n_R(k) \cdot n_S(k))$. Based on this observation, we tested {\tt NaturalJoin} on five different datasets. 
        \begin{enumerate}
            \item The first three data sets are such that $S$ is not skewed (for each $k$, $n_S(k) = \Theta(1)$) while $R$ is generated following a Zipf distribution~\cite{zipf} with skewness $1.0$, $1.5$, and $2.0$, respectively;
            \item Both $R$ and $S$ follow a Zipf distribution, with skewness $2.0$ in $R$ and $1.0$ in $S$;
            \item Both $R$ and $S$ are generated following a Zipf distribution with skewness $1.5$.
        \end{enumerate}
    Since $n_S(k) = \Theta(1)$, the running time of the reduce function on the first three datasets becomes $\Theta(n_R(k))$, i.e., linear. It remains superlinear in the last two datasets.

\smallskip
\noindent $\bullet$~~{\tt MatMult} is taken from a MapReduce library for sparse matrix multiplication~\cite{CS15} and implements a blocked matrix multiplication algorithm. The map function duplicates each block as many times as the number of products for which the block is required. The reduce function multiplies two input blocks with running time $\Theta(n^3 \cdot d \cdot d')$, where $n$ is the block side, and $d$ and $d'$ are the block densities. We considered three variants of {\tt MatMult}, using different strategies for partitioning keys:
        \begin{itemize}
            \item In {\tt MatMult-Opt} the number of block products assigned to each reduce task is optimally balanced, as described in~\cite{CS15} (this is the default library partitioner).
            \item In {\tt MatMult-Rnd} block products are assigned randomly among reduce tasks. This may result in some reduce data unbalancing.
            \item In {\tt MatMult-Unbal} we forced a very unbalanced partition. Let $k$ be the number of matrix blocks. Among the $k^{3/2}$ block products performed by the algorithm, the most computationally demanding $k$ products are assigned to a single reduce task, while the remaining ones are fairly distributed. 
        \end{itemize}
We have tested each {\tt MatMult} variant on two input datasets. In the former, matrix items are uniformly distributed and all matrix blocks have the same expected density $d = d' = 0.25$. In the latter, block densities are different depending on the block position: blocks $(i, j)$ are such that $d = 1 / 2^{\sqrt{k} - 1 - j}$ and $d' = 1 / 2^{\sqrt{k} - 1 - i}$, where $k$ is the number of blocks in each matrix. Intuitively, density increases across columns or rows in the two input matrices.

\begin{figure*}[t]
\begin{center}
\begin{tabular}{ccc}
$~~~~~~~~~~~~~~~~~~~~~~~~~~~~~~~~~~$ & 
$~~~~~~~~~~~~~~~~~~~~~~~~~~~~~$$~~~~~~~~~~~~~~~~~~~~~~~~~~~~~~~~~~~~~~~~~$ & 
\includegraphics[width=0.9\columnwidth]{charts/plots/legend}$~~~~~$ \vspace{-6mm} \\
\end{tabular}
\begin{tabular}{cccc}
\hspace{-3mm}\includegraphics[width=0.40\columnwidth]{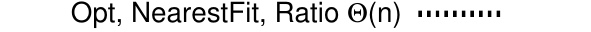} & & & \vspace{-1mm}\\
\hspace{-3mm}\includegraphics[width=0.20\columnwidth]{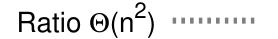}$~~~~$
\hspace{-3mm}\includegraphics[width=0.12\columnwidth]{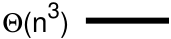}$~~~~~$
\hspace{-3mm}\includegraphics[width=0.12\columnwidth]{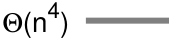}
& & & \vspace{-7mm}\\
\hspace{-3mm}\includegraphics[width=0.50\columnwidth]{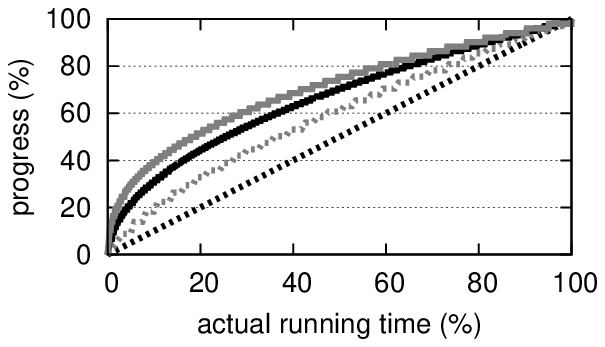} & 
\hspace{-3mm}\includegraphics[width=0.50\columnwidth]{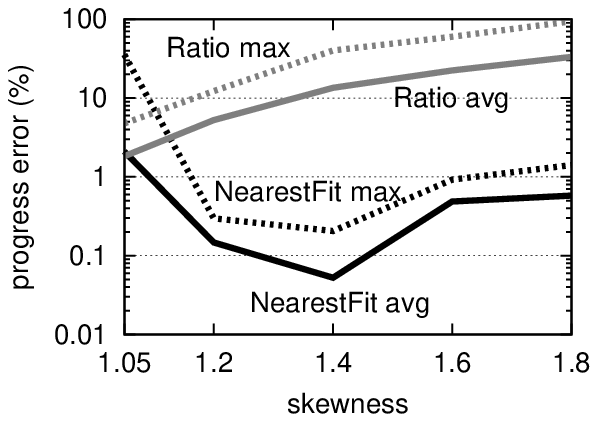} &
\hspace{-3mm}\includegraphics[width=0.50\columnwidth]{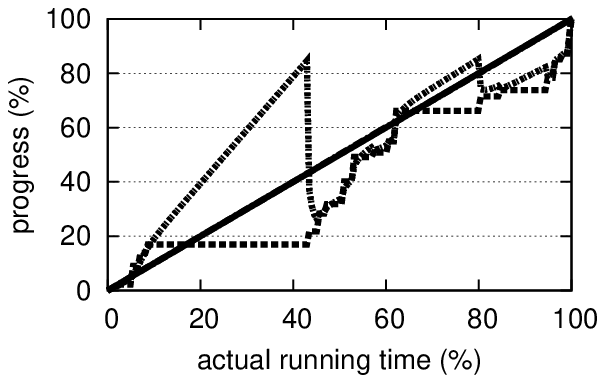} &
\hspace{-3mm}\includegraphics[width=0.50\columnwidth]{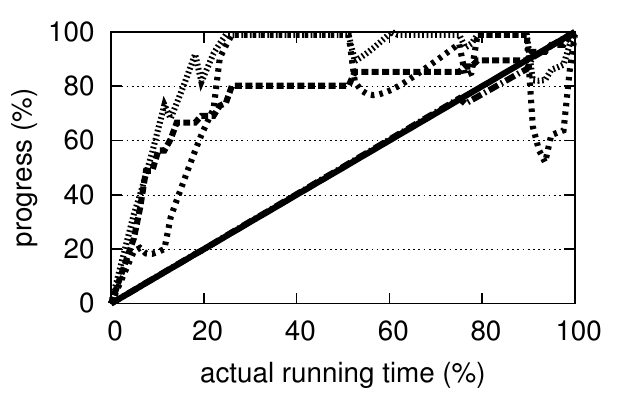} \\ 
(a) & (b) & (c) & (d) \\ 
\end{tabular}
\vspace{-3mm}
\end{center}
\caption{Experiments on the synthetic benchmark: (a) progress plots for different reduce running times; (b) error of progress estimates for $\Theta(n^2)$ reduce functions on datasets with increasing skewness; (c--d) progress plots obtained with 1 and 8 reduce tasks, respectively.}
\label{fi:sythentic}
\end{figure*}

\subsection{Metrics}


In our experimental analysis we assessed prediction accuracy, slowdown, and space overhead of {\tt NearestFit}.

\paragraph{Progress accuracy.} This is obtained by computing the absolute percentage error between the estimated and the optimal progress:
\[ error(t) = \left | \frac{t - t_{start}}{ \widetilde{e}(t) - t_{start} } - \frac{t - t_{start}}{ e(t) - t_{start} } \right | \times 100\]
Let $T$ be the set of prediction times, i.e., times at which progress is updated. Then the mean and maximum errors are computed as follows:

\vspace{2mm}

\noindent\begin{tabular}{ lr }
  $avgErr = \frac{1}{|T|} \sum_{t\in T} error(t)$ & $~~~maxErr = \max\limits_{t\in T}  error(t)$ 
\end{tabular}

\paragraph{Slowdown.} We compare the native running times with executions under {\tt NearestFit}. 

\paragraph{Space overhead.} Let $|\text{\em map profile}|$ and $|\text{\em reduce profile}|$ be the cumulative sizes for profiles collected among all map tasks and all reduce tasks, respectively. To evaluate space overhead, we compare the amount $|\text{\em map profile}|+|\text{\em reduce profile}|$ with the shuffle data size:
\[ overhead = \frac{|\text{\em map profile}| + |\text{\em reduce profile}|}{|\text{\em shuffle data}|} \times 100\]

\subsection{Platform and Hadoop configuration}

The experiments have been carried out on three different Amazon EC2 clusters, running a customized release of Hadoop 2.6.0. Besides a node for resource managers, the three clusters included 8, 16, and 32 workers devoted to both Hadoop tasks and the HDFS. We used Amazon EC2 {\tt m1.xlarge} instances, each providing 4 virtual cores, 15 GiB of main memory, and 840 GiB of secondary storage. 
Since one worker runs the MapReduce application master,  the actual parallelism available for map and reduce tasks is decreased by one on each cluster. We have considered only clusters composed by homogeneous machines and disabled speculative execution.
 
Among several Hadoop runtime parameters, the number $R$ of reduce tasks plays a crucial role during the execution of a job. If $R$ is set to the (actual) task parallelism available on a cluster, we have a single wave execution where all reduce tasks can be immediately started after the map phase. 
If $R$ is larger than the available cluster parallelism, then the amount of shuffle data can be partitioned among a higher number of tasks, decreasing the task input size and possibly reducing the impact of straggling instances. Multiple wave executions, however, can increase the framework overhead. We have thus considered two execution scenarios:
\begin{itemize}
    \item {\em single-wave}, setting $R$ to $7$, $15$, and $31$ on the three clusters;
    \item {\em multiple-waves}, setting $R$ to two times the cluster parallelism (i.e., 14, 30, and 62, respectively).
\end{itemize}
This choice has been driven by the optimization suggestions provided by the official Hadoop documentation~\cite{hadoop}.


\section{Experimental evaluation}
\label{se:experiments}

In this section we discuss the outcome of an extensive empirical evaluation, which required roughly 500 cluster hours over the EC2 platform. This is a very optimistic estimate: it does not include times for cluster configuration, setup times for each single experiment, collection and processing of experimental data, debugging issues, and tuning of progress indicators (e.g.,  {\tt NearestFit} experiments with different threshold choices). The description focuses on the reduce phase. As observed in Section~\ref{ss:model-generalization}, we used a linear model in the map phase and results across all benchmarks consistently proved to be very accurate.

\subsection{A warm-up example}

As a preliminary experiment, we used our synthetic benchmark to study the interplay between data skewness and time complexity of the reduce functions. We fed different versions of the synthetic benchmark (linear, quadratic, cubic, and quartic) with data sets characterized by different skewness levels. The outcome of the experiment is summarized in Figure~\ref{fi:sythentic} and confirms our hypothesis that linear progress models are harmed by data skewness and large running times. 

In Figure~\ref{fi:sythentic}a, the dataset is mildly skewed  and only the time complexity changes. In the $\Theta(n)$ case, all progress indicators roughly match the optimal progress. As the running time grows, differently from {\tt NearestFit}, {\tt Ratio} becomes more and more inaccurate (the job and task versions, as well as {\tt Hadoop}, overlap and we plot only one of them).

In Figure~\ref{fi:sythentic}b we performed the symmetric experiment, fixing the time complexity to $\Theta(n^2)$ and increasing the data skewness $\sigma$ from $1.05$ (almost unskewed) to $1.8$ (rather skewed). Average and maximum error of {\tt Ratio} increase with $\sigma$, while the maximum error of {\tt NearestFit} is worse for unskewed data. Notice the log-scale on the $y$-axis. Overall, {\tt NearestFit} appears to be much more accurate than {\tt Ratio} and its error is not badly affected by skewness: the average error of {\tt NearestFit} stays below $1\%$, while can be as much as $33\%$ for {\tt Ratio}.

In the previous experiments we used a single reduce task: Figure~\ref{fi:sythentic}c shows the progress plot for $\sigma=1.4$ and quadratic complexity. The error increases with the number of tasks, especially for {\tt Ratio} and {\tt Hadoop}, as shown by the progress plot in Figure~\ref{fi:sythentic}d (same parameters, 8 reduce tasks).

\begin{table*}
\begin{scriptsize}
\begin{center}
\begin{tabular}{|l||c|c|c|c|c||c|c|c|c|}\hline

\multicolumn{1}{|c||}{} &
\multirow{2}{*}{\sc Dataset} &
\multicolumn{4}{|c||}{\sc Mean  Absolute Percentage Error} &
\multicolumn{4}{|c|}{\sc Max  Absolute Percentage Error} \\\cline{3-10}

\multicolumn{1}{|c||}{} &
\multicolumn{1}{|c|}{} &
\multicolumn{1}{|c|}{\sc Hadoop} &
\multicolumn{1}{|c|}{\sc JobRatio} &
\multicolumn{1}{|c|}{\sc TaskRatio} &
\multicolumn{1}{|c||}{\sc NearestFit} &

\multicolumn{1}{|c|}{\sc Hadoop} &
\multicolumn{1}{|c|}{\sc JobRatio} &
\multicolumn{1}{|c|}{\sc TaskRatio} &
\multicolumn{1}{|c|}{\sc NearestFit} \\\hline\hline

\texttt{WordCount} & {\tt wiki} & 11.30 & 13.02 & 13.01 & 12.44 & 19.02 & 21.34 & 21.34 & 20.80 \\\hline
\texttt{WordCount-NC} & {\tt wiki} & 6.68 & 0.48 & 0.53 & 0.95 & 16.27 & 1.56 & 1.56 & 3.10 \\\hline

\multirow{2}{*}{\tt InvertedIndex} & {\tt wiki 5K} & 6.85 & 4.11 & 5.88 & 8.03 & 13.46 & 12.86 & 17.65 & 18.59 \\
\multicolumn{1}{|c||}{} & {\tt wiki 50K} & 7.94 & 12.34 & 20.16 & 1.76 & 17.66 & 21.05 & 37.31 & 5.51 \\\hline

\multirow{6}{*}{\tt 2PathGenerator} & {\tt loc-Gowalla} & 10.91 & 39.47 & 36.30 & 4.90 & 34.23 & 87.51 & 87.51 & 12.34 \\
\multicolumn{1}{|c||}{} & {\tt web-Google} & 3.04 & 9.40 & 7.39 & 0.76 & 5.53 & 16.02 & 14.94 & 1.73 \\
\multicolumn{1}{|c||}{} & {\tt web-Stanford} & 14.40 & 25.35 & 13.86 & 1.19 & 36.53 & 48.29 & 24.60 & 2.95 \\
\multicolumn{1}{|c||}{} & {\tt com-Youtube} & 23.12 & 31.78 & 18.24 & 3.09 & 48.20 & 58.29 & 46.16 & 7.15 \\
\multicolumn{1}{|c||}{} & {\tt web-Berkstan} & 27.51 & 31.84 & 22.29 & 2.40 & 51.81 & 92.29 & 91.85 & 5.75 \\
\multicolumn{1}{|c||}{} & {\tt as-Skitter} & 26.21 & 26.60 & 21.46 & 1.07 & 46.30 & 66.05 & 60.51 & 2.33 \\\hline

\multirow{6}{*}{\tt TriangleCount} & {\tt loc-Gowalla} & 7.18 & 0.47 & 0.66 & 9.24 & 11.74 & 1.54 & 2.07 & 35.23 \\
\multicolumn{1}{|c||}{} & {\tt web-Google} & 1.81 & 1.10 & 0.53 & 0.22 & 4.50 & 1.78 & 0.79 & 0.45 \\
\multicolumn{1}{|c||}{} & {\tt web-Stanford} & 1.92 & 0.57 & 0.17 & 0.10 & 3.98 & 0.84 & 0.28 & 0.29 \\
\multicolumn{1}{|c||}{} & {\tt com-Youtube} & 1.55 & 0.96 & 1.00 & 0.90 & 4.34 & 1.92 & 2.15 & 2.01 \\
\multicolumn{1}{|c||}{} & {\tt web-Berkstan} & 3.30 & 0.85 & 1.58 & 1.44 & 8.77 & 2.25 & 4.49 & 4.37 \\
\multicolumn{1}{|c||}{} & {\tt as-Skitter} & 4.43 & 2.78 & 2.68 & 4.65 & 7.02 & 6.19 & 5.94 & 7.97 \\\hline

\multirow{5}{*}{\tt NaturalJoin} & {\tt linear} $1.0$ & 6.53 & 3.91 & 1.46 & 1.40 & 16.41 & 7.02 & 2.81 & 9.58 \\
\multicolumn{1}{|c||}{} & {\tt linear} $1.5$ & 25.85 & 1.12 & 1.14 & 2.34 & 48.55 & 1.81 & 1.81 & 5.02 \\
\multicolumn{1}{|c||}{} & {\tt linear} $2.0$ & 30.24 & 3.55 & 3.56 & 0.76 & 58.91 & 7.37 & 7.37 & 2.72 \\
\multicolumn{1}{|c||}{} & {\tt sl} $2.0\,1.0$ & 31.81 & 45.49 & 45.62 & 11.11 & 70.17 & 95.96 & 95.96 & 22.77 \\
\multicolumn{1}{|c||}{} & {\tt sl} $1.5$ & 34.10 & 48.14 & 48.24 & 1.68 & 72.84 & 97.37 & 97.37 & 3.73 \\\hline

\multirow{2}{*}{\tt MatMult-Opt} & {\tt skewed} & 5.18 & 11.16 & 10.81 & 1.43 & 9.88 & 21.59 & 20.66 & 7.41 \\
\multicolumn{1}{|c||}{} & {\tt uniform} & 7.01 & 1.50 & 1.36 & 0.47 & 12.46 & 3.10 & 2.84 & 0.87 \\\hline

\multirow{2}{*}{\tt MatMult-Rnd} & {\tt skewed} & 12.51 & 11.94 & 11.23 & 0.71 & 22.56 & 22.26 & 21.16 & 3.97 \\
\multicolumn{1}{|c||}{} & {\tt uniform} & 5.75 & 0.81 & 0.83 & 0.30 & 12.38 & 1.64 & 1.70 & 0.49 \\\hline

\multirow{2}{*}{\tt MatMult-Unbal} & {\tt skewed} & 38.20 & 9.09 & 4.65 & 0.14 & 77.06 & 15.04 & 12.05 & 2.08 \\
\multicolumn{1}{|c||}{} & {\tt uniform} & 14.92 & 0.61 & 0.72 & 0.35 & 37.91 & 1.31 & 1.62 & 1.07 \\\hline

\hline\hline

{\sc Arithm. mean} & & 13.71 & 12.53 & 10.94 & 2.73 & 28.46 & 26.45 & 25.35 & 7.05 \\

\hline

\end{tabular}

\end{center}
\end{scriptsize}
\vspace{-2mm}\nocaptionrule\caption{Accuracy of progress indicators with 8 workers and single wave execution.}
\label{ta:progress-table}
\end{table*}

\subsection{Progress indicator accuracy}


In this section we compare the accuracy of {\tt NearestFit} with the state-of-the-art progress indicators on the real benchmarks. The main outcome of our analysis is summarized in Table~\ref{ta:progress-table}, which reports the mean and maximum absolute percentage errors across all benchmarks and data sets, computed on the smallest cluster. The arithmetic mean of the errors is shown on the last line, and gives a clear clue on the accuracy of the different indicators. The scenario is quite interesting and diversified if we examine each specific benchmark.

{\tt NearestFit} is especially good at predicting progress for high time complexity and high skewness. The emblematic example is {\tt 2PathGenerator}, where both {\tt Hadoop} and {\tt Ratio} exhibit very poor accuracy. For instance, the maximum error of {\tt JobRatio} can be as large as $92\%$. And there is not just a single wrong prediction across each execution, but predictions appear to be repeatedly incorrect, as proved by the high values of the average error. {\tt TaskRatio} is slightly better than {\tt JobRatio}, though the two approaches are overall quite similar.

\begin{figure}[t]
\begin{center}
\includegraphics[width=0.99\columnwidth]{charts/plots/legend}
\begin{tabular}{cc}
\hspace{-3mm}\includegraphics[width=0.50\columnwidth]{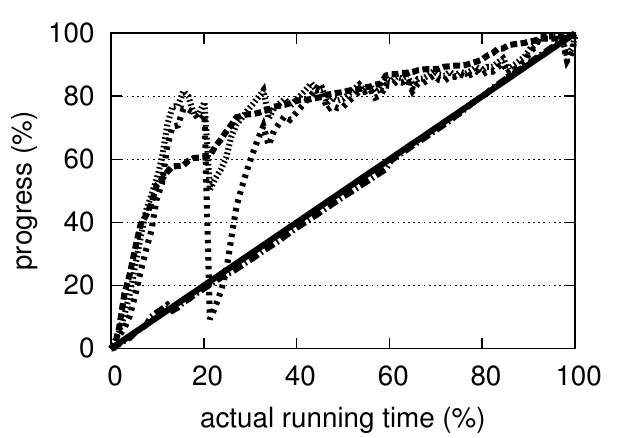} &
\hspace{-3mm}\includegraphics[width=0.50\columnwidth]{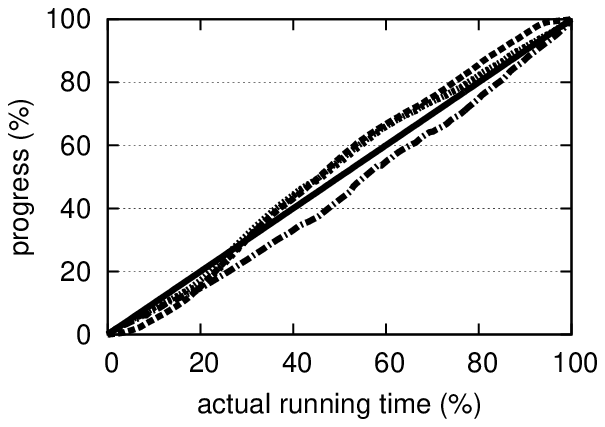} \\
\end{tabular}
\end{center}
\vspace{-3.5mm}
\caption{Progress prediction on a quadratic benchmark ({\tt 2PathGenerator}, left column) and a linear benchmark ({\tt TriangleCount}, right column) on the smallest cluster and dataset {\tt as-Skitter}.}
\label{fi:linear-quadratic}
\end{figure}

On benchmarks for which reduce functions are less computationally demanding (linear time), both {\tt NearestFit} and {\tt Ratio} are accurate: the latter is sometimes better (see, e.g., {\tt TriangleCount} on datasets {\tt as-Skitter} or {\tt loc-Gowalla}), but the average error of {\tt NearestFit} is always very reasonable. The maximum error has a peak on {\tt loc-Gowalla}, whose execution time is however quite short: the reduce phase takes less than 4 minutes. The different behavior of {\tt Ratio} on the quadratic benchmark  {\tt 2PathGenerator} and the linear benchmark {\tt TriangleCount} is exemplified by Figure~\ref{fi:linear-quadratic}, which shows the progress plots obtained for dataset {\tt as-Skitter}. It is worth noticing that not only the error of {\tt Ratio} is significant on benchmark {\tt 2PathGenerator}, but there are also wide prediction fluctuations, ranging from underestimates to large overestimates.

\begin{figure}[t]
\begin{center}
\includegraphics[width=0.99\columnwidth]{charts/plots/legend}
\begin{tabular}{cc}
\hspace{-3mm}\includegraphics[width=0.50\columnwidth]{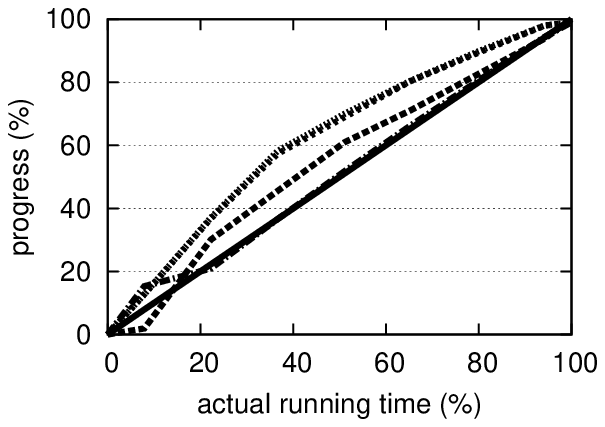} &
\hspace{-3mm}\includegraphics[width=0.50\columnwidth]{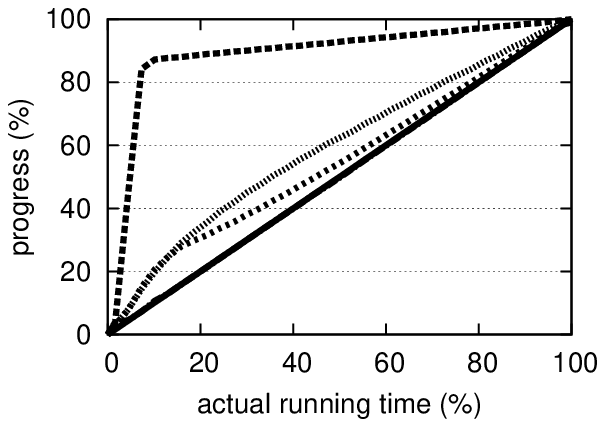} \\
\hspace{-3mm}\includegraphics[width=0.50\columnwidth]{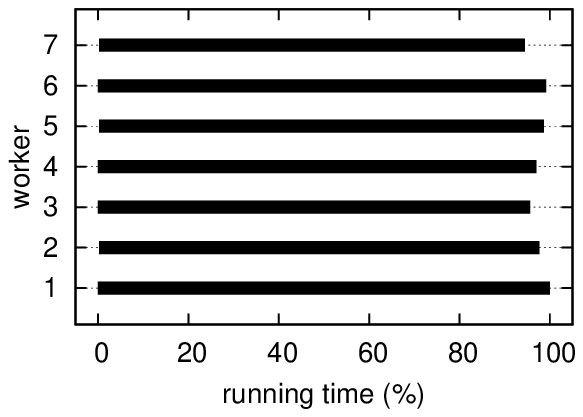} &
\hspace{-3mm}\includegraphics[width=0.50\columnwidth]{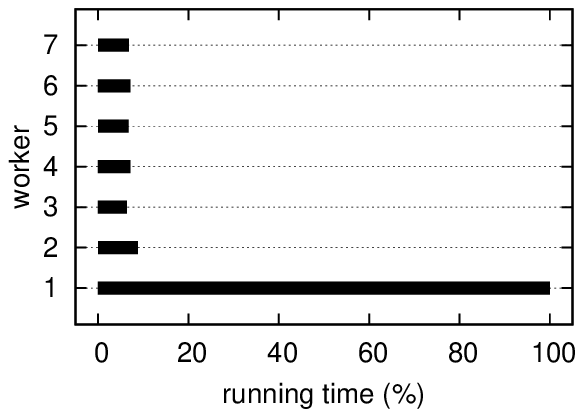} \\
\end{tabular}
\end{center}
\vspace{-3.5mm}
\caption{Load balancing issues: {\tt MatrixMult-Opt} and {\tt MatrixMult-Unbal} for skewed matrices on the smallest cluster (progress plots and swimlanes plots).}
\label{fi:matrix}
\end{figure}

\begin{figure}[t]
\begin{center}
\includegraphics[width=0.90\columnwidth]{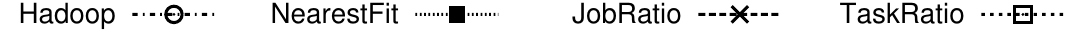}
\begin{tabular}{cc}
\hspace{-3mm}\includegraphics[width=0.50\columnwidth]{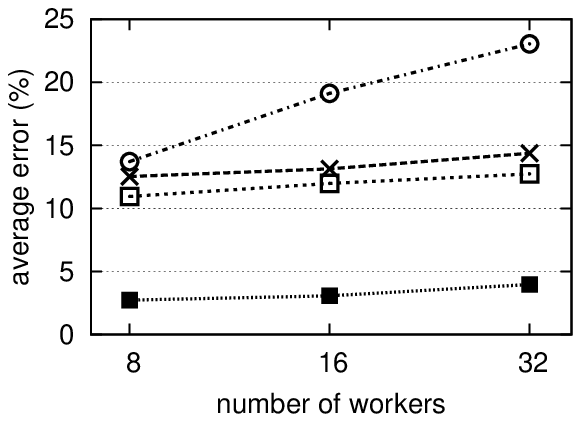} & 
\hspace{-3mm}\includegraphics[width=0.50\columnwidth]{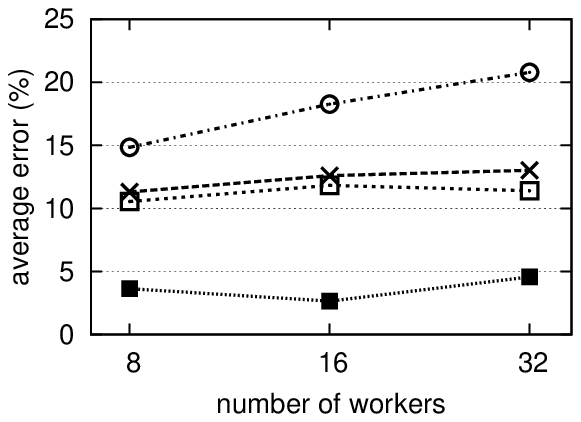} \\ 
\end{tabular}
\end{center}
\vspace{-3mm}
\caption{Accuracy of the progress indicators on different clusters for a single wave (left) or multiple waves (right) of reduce tasks.}
\label{fi:progress-accuracy}
\end{figure}

The {\tt Hadoop} progress indicator is instead mainly affected by load balancing issues between tasks:  load unbalancing can yield very wrong estimates, as exemplified for benchmarks {\tt MatMult-Opt} and  {\tt MatMult-Unbal} in Figure~\ref{fi:matrix}: while {\tt NearestFit} is close to optimal and {\tt Ratio} is reasonably accurate in both experiments,  {\tt Hadoop} predictions are very poor with the unbalanced partitioner. This is because the progress of short tasks becomes quickly $100\%$ (short tasks complete early, as shown in the swimlanes plot),  yielding a large average progress for the entire phase. 

The error of {\tt Hadoop} increases on larger clusters, where the impact of the many short tasks on progress analysis becomes more and more noticeable. This is confirmed by Figure~\ref{fi:progress-accuracy}, which plots the mean of the average errors across all benchmarks and datasets on three different clusters. {\tt Ratio} and {\tt NearestFit} are only slightly affected by larger degrees of parallelism, and the results of Table~\ref{ta:progress-table} are confirmed on all clusters, both for single and for multiple waves executions.

\subsection{Slowdown}

\begin{figure}[t]
\hspace{11mm}\includegraphics[width=0.35\columnwidth]{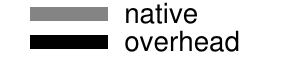}
\vspace{-13mm}
\begin{center}
\includegraphics[width=1.0\columnwidth]{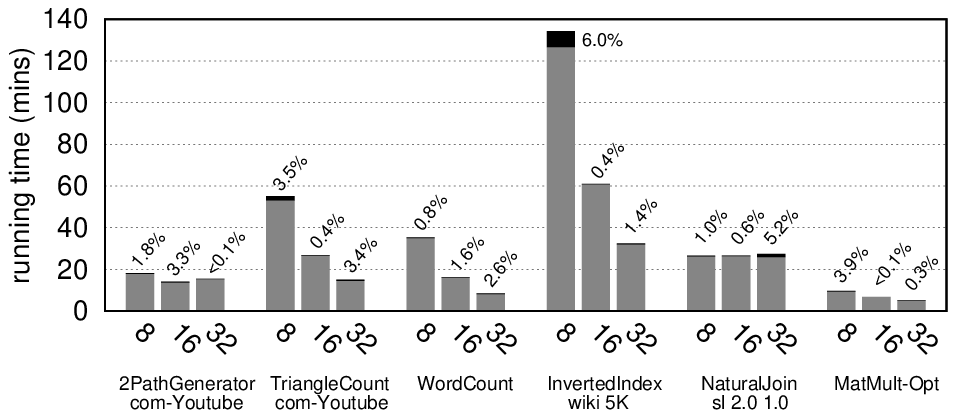}
\end{center}
\vspace{-2mm}
\caption{Slowdown with respect to native execution on a selection of representative benchmarks on three different clusters.}
\label{fi:slowdown-bar}
\end{figure}

\begin{figure}[t]
\begin{center}
\begin{tabular}{cc}
\hspace{-3mm}\includegraphics[width=0.50\columnwidth]{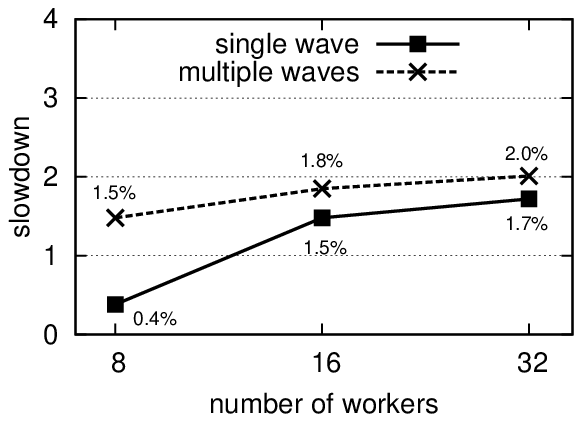} &
\hspace{-3mm}\includegraphics[width=0.50\columnwidth]{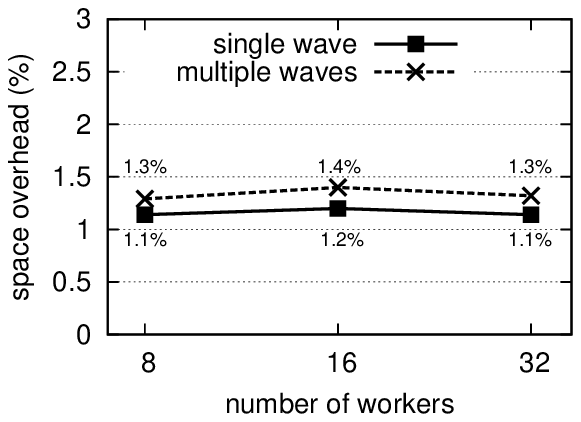} \vspace{-1mm}\\
(a) & (b) \\
\end{tabular}
\end{center}
\vspace{-3.5mm}
\caption{Average slowdown and space overhead for executions with single and multiples waves of reduce tasks.}
\label{fi:slowdown-space}
\end{figure}

We now discuss the overhead introduced by {\tt NearestFit} on the wall-clock time of native executions. Performance figures on a selection of representative benchmarks are shown in Figure~\ref{fi:slowdown-bar}, which reports for each benchmark the slowdown on three different clusters. 
On linear benchmarks and unskewed data, using larger clusters yields shorter running times. This is not the case  on superlinear benchmarks and skewed data, where a larger degree of parallelism cannot be fully exploited due to stragglers: the running time in {\tt 2PathGenerator} and {\tt NaturalJoin}, for instance, stays almost constant. The slowdown is always small: no execution was more than $1.06\times$ w.r.t. native execution. The average slowdown across all benchmarks, including those not reported in Figure~\ref{fi:slowdown-bar}, was $0.4\%$, as shown in Figure~\ref{fi:slowdown-space}a (8 workers, single wave). 
Figure~\ref{fi:slowdown-space}a also shows that the average slowdown slightly increases on larger clusters: this is expected on average, as native executions can be shorter while the collected profile data stays the same.  


\subsection{Space overhead}

We first focus on the profile size, which determines communication costs. As shown in Figure~\ref{fi:slowdown-space}b, the space overhead is negligible and independent of the cluster size. The space usage is small because on the map side information on implicit keys is aggregated and on the reduce side we use smoothing (see Section~\ref{ss:gathering-opt} and Section~\ref{se:implementation}, respectively).  Cluster independence follows from the observation that the number of map tasks depends only on the job input size. The number of reduce tasks, on the other side, changes on different clusters, but the number of spawned reduce functions is exactly the same: small -- unpredictable -- variabilities of the size of reduce task profiles are due to different scheduling strategies, that might yield different bursts (see Section~\ref{se:implementation}). Benchmark-specific figures on the small cluster are given in Table~\ref{ta:space-table}.

\begin{table*}[t]
\vspace{1mm} 
\begin{scriptsize}
\begin{center}
\begin{tabular}{|l||c|c|c|c|c|c|c|}\hline

\multicolumn{1}{|c||}{} &
{\sc Dataset}  &
\multicolumn{1}{|c|}{\sc input (mb)} &
\multicolumn{1}{|c|}{\sc shuffle data (mb)} &
\multicolumn{1}{|c|}{\sc output (mb)} &
\multicolumn{1}{|c|}{\sc map profile (mb)} &
\multicolumn{1}{|c|}{\sc reduce profile (mb)} &
\multicolumn{1}{|c|}{\sc overhead (\%)} \\\hline\hline

\texttt{WordCount} & {\tt wiki} & 50202.70	& 9812.04 & 5600.17 & 20.01 & 2.58 & 0.23 \\\hline
\texttt{WordCount-NC} & {\tt wiki} & 50202.70 & 21715.84 & 891.43 & 21.12 & 1.67 & 0.10 \\\hline

\multirow{2}{*}{\tt InvertedIndex} & {\tt wiki 5K} & 51112.74 & 12962.94 & 12784.75 & 265.93 & 5.25 & 2.09\\
\multicolumn{1}{|c||}{} & {\tt wiki 50K} & 51112.74 & 18277.93 & 18019.26 & 1708.35 & 7.62 & 9.39 \\\hline

\multirow{6}{*}{\tt 2PathGenerator} & {\tt loc-Gowalla} & 32.03 & 35.66	& 4395.35 & 0.38 & 0.98 & 3.82\\
\multicolumn{1}{|c||}{} & {\tt web-Google} & 161.61 & 178.09 & 14322.27 & 0.38 & 4.31 & 2.64 \\
\multicolumn{1}{|c||}{} & {\tt web-Stanford} & 75.46 & 83.06 & 74930.03 & 0.38 & 2.07 & 2.96 \\
\multicolumn{1}{|c||}{} & {\tt com-Youtube} & 108.35 & 119.75 & 25686.61 & 0.39 & 2.33 & 2.27\\
\multicolumn{1}{|c||}{} & {\tt web-Berkstan} & 259.96 & 285.33 & 501082.86 & 0.78 & 4.79 & 1.95  \\
\multicolumn{1}{|c||}{} & {\tt as-Skitter} & 425.05 & 467.38 & 287677.55 & 0.39 & 8.95 & 2.00 \\\hline

\multirow{6}{*}{\tt TriangleCount} & {\tt loc-Gowalla} & 4431.45 & 4977.59 & 9.74 & 4.05 & 4.94 & 0.18 \\
\multicolumn{1}{|c||}{} & {\tt web-Google} & 14497.59 & 15856.18 & 59.39 & 12.04 & 12.66 & 0.16\\
\multicolumn{1}{|c||}{} & {\tt web-Stanford} & 75078.55 & 82518.25 & 28.30 & 59.99 & 32.88 & 0.11\\
\multicolumn{1}{|c||}{} & {\tt com-Youtube} & 25819.82 & 28595.63 & 19.47 & 21.10 & 17.40 & 0.13 \\
\multicolumn{1}{|c||}{} & {\tt web-Berkstan} & 501832.44 & 554678.13 & 98.44 & 399.34 & 50.16 & 0.08 \\
\multicolumn{1}{|c||}{} & {\tt as-Skitter} & 288383.69 & 318605.68 & 130.04 & 229.31 & 159.68 & 0.12\\\hline

\multirow{5}{*}{\tt NaturalJoin} & {\tt linear} $1.0$ & 4283.23 & 5,822.30 & 83,924.16 & 2.00 & 1.69 & 6.33E-02  \\
\multicolumn{1}{|c||}{} & {\tt linear} $1.5$ & 4258.59 & 5797.72 & 83923.36 & 1.89 & 0.73 & 4.52E-02 \\
\multicolumn{1}{|c||}{} & {\tt linear} $2.0$ & 4258.59 & 5797.72 & 83923.36 & 1.89 & 0.73 & 4.52E-02 \\
\multicolumn{1}{|c||}{} & {\tt sl} $2.0\,1.0$ & 8.44 & 11.49 & 50758.64 & 0.07 & 0.08 & 1.34 \\
\multicolumn{1}{|c||}{} & {\tt sl} $1.5$ & 8.43 & 11.48 & 148455.27 & 0.11 & 0.02 & 1.14\\\hline

\multirow{2}{*}{\tt MatMult-Opt} & {\tt skewed} & 1021.96 & 8907.43 & 14010.68 & 0.02 & 0.01 & 4.18E-04 \\
\multicolumn{1}{|c||}{} & {\tt uniform} & 933.84 & 7856.49 & 15703.94 & 0.02 & 0.01 & 4.53E-04 \\\hline

\multirow{2}{*}{\tt MatMult-Rnd} & {\tt skewed} & 1021.96 & 8907.43 & 14010.68 & 0.02 & 0.01 & 4.18E-04 \\
\multicolumn{1}{|c||}{} & {\tt uniform} & 933.84 & 7856.49 & 15,703.94 & 0.02 & 0.01 & 4.50E-04\\\hline

\multirow{2}{*}{\tt MatMult-Unbal} & {\tt skewed} & 1021.96 & 8907.43 & 14010.68 & 0.02 & 0.01 & 4.21E-04\\
\multicolumn{1}{|c||}{} & {\tt uniform} & 933.84 & 7856.49 & 15703.94 & 0.02 & 0.01 & 4.51E-04 \\


\hline

\end{tabular}

\end{center}
\end{scriptsize}
\vspace{-2mm}\nocaptionrule\caption{Exact figures for the space overhead of {\tt NearestFit} (8 workers, single wave).}
\label{ta:space-table}
\end{table*}

We also analyzed the memory peak on worker nodes and on the application master. We omit detailed results since it turned out not to be a bottleneck. Memory usage in our implementation is indeed limited by fixed constants: $\lambda$ on the worker side, and space-efficient streaming data structures on the master side (see Section~\ref{ss:gathering-opt}).

\subsection{Time/space/accuracy tradeoffs}

In Section~\ref{se:practical} we introduced a distinction between implicit and explicit keys and we observed that this is crucial to make {\tt NearestFit} practical. We now back up this observation with experimental data.

\begin{figure*}[t]
\begin{center}
\begin{tabular}{ccccc}
\hspace{-3mm}\includegraphics[width=0.4\columnwidth]{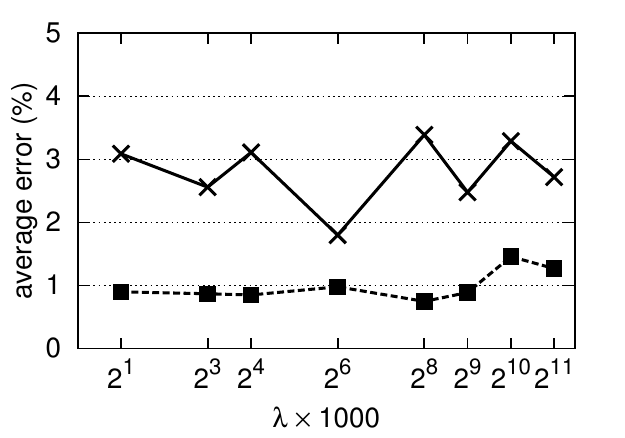} & 
\hspace{-3mm}\includegraphics[width=0.4\columnwidth]{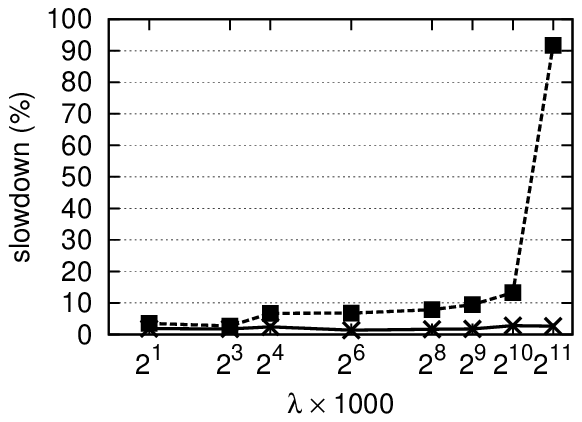} &
\hspace{-3mm}\includegraphics[width=0.4\columnwidth]{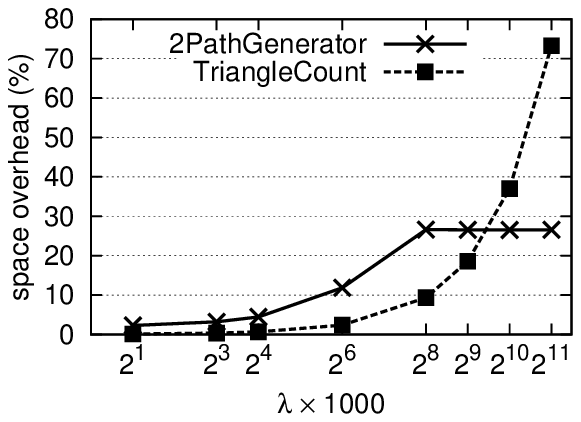} & 
\hspace{-3mm}\includegraphics[width=0.4\columnwidth]{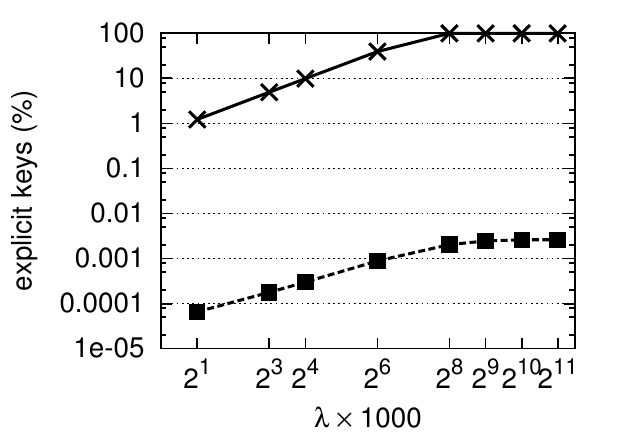} &
\hspace{-3mm}\includegraphics[width=0.4\columnwidth]{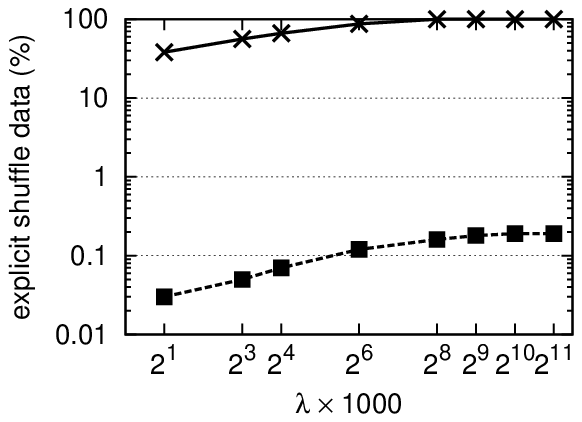} \\ 
(a) & (b) & (c) & (d) & (e)
\vspace{-3mm}
\end{tabular}
\end{center}
\caption{Time/space/accuracy tradeoffs with different numbers of explicit keys: (a) accuracy; (b) slowdown; (c) space overhead; (d) percentage of explicit keys; (e) percentage of the shuffle data associated to explicit keys.}
\label{fi:tradeoff}
\end{figure*}

Figure~\ref{fi:tradeoff} considers the {\tt 2PathGenerator} and {\tt TriangleCount} benchmarks on the {\tt com-Youtube} dataset. We run {\tt NearestFit} by exponentially increasing the number $\lambda$ of explicit keys (in all the experiments discussed so far $\lambda=2000$). When increasing $\lambda$, we also change the size of the streaming data structure used by the application master to merge explicit keys: this is set to $35\lambda$, which is much smaller than the total number of explicit keys collected across all the $410$ map tasks of {\tt com-Youtube}. 

Figure~\ref{fi:tradeoff}a shows that accuracy does not benefit of larger values of $\lambda$. On the other hand, as shown in Figure~\ref{fi:tradeoff}b, the largest $\lambda$ values yield a steep increase of the running times, due to garbage collection and  communication costs. Space usage is also harmed by $\lambda$, since map task profiles become larger as $\lambda$ increases. On the {\tt 2PathGenerator}, the overhead flattens at $\lambda=2^8\times 1000$ because the number of distinct keys is smaller than this value.

Figure~\ref{fi:tradeoff}d plots the percentage of explicit keys after merging, with respect to the total number of distinct keys. When $\lambda\geq 2^8\times 1000$, all the keys in {\tt 2PathGenerator} are explicit as already observed. There is a remarkable difference between the two benchmarks: in {\tt TriangleCount}, the percentage of explicit keys is much smaller than {\tt 2PathGenerator} (notice the log scale on the $y$-axis). This is because map functions  in {\tt TriangleCount} produce a large number of very small key groups. The flat trend for $\lambda\geq 2^8\times 1000$ therefore depends on the streaming data structure. Figure~\ref{fi:tradeoff}e shows how much shuffle data is associated with explicit keys: a comparison of Figures~\ref{fi:tradeoff}d and~\ref{fi:tradeoff}e reveals that, even for $\lambda=2000$, $1\%$ of keys  is associated with $38\%$ of shuffle data in {\tt 2PathGenerator}, due to data skewness. This suggests that a small space is sufficient to characterize the heaviest keys without compromising accuracy.


\subsection{Nearest neighbor regression vs. curve fitting}

\begin{figure}[t]
\begin{center}
\begin{tabular}{cc}
\hspace{0mm}\includegraphics[width=0.45\columnwidth]{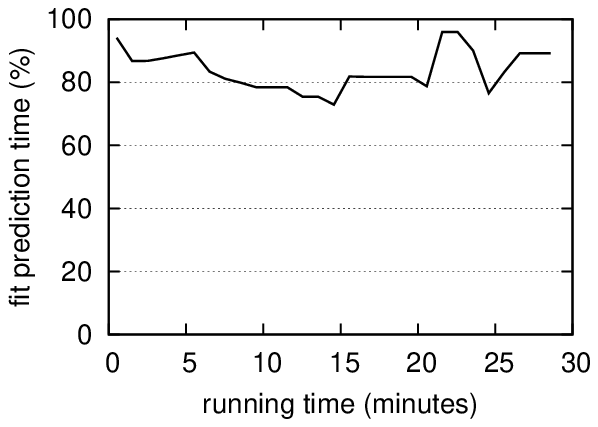} & 
\hspace{-0mm}\includegraphics[width=0.45\columnwidth]{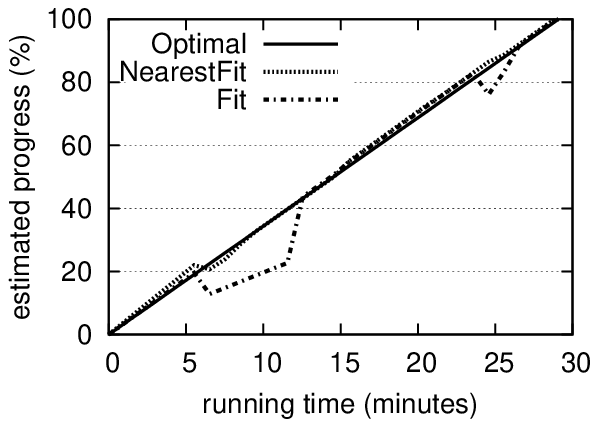} \\
(a) & (b) \\ 
\hspace{-0mm}\includegraphics[width=0.45\columnwidth]{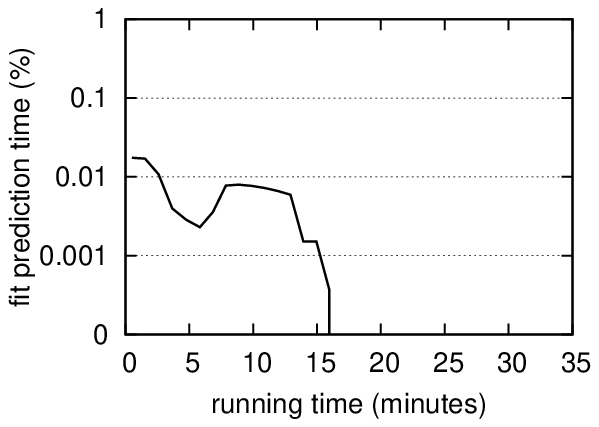} & 
\hspace{-0mm}\includegraphics[width=0.45\columnwidth]{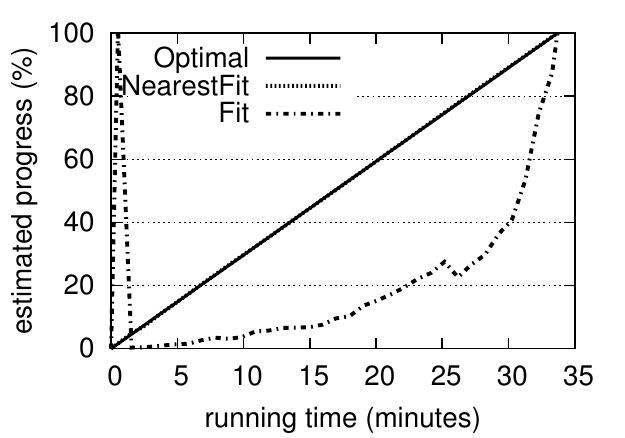}  \\
(c) & (d) 
\vspace{-3mm}
\end{tabular}
\end{center}
\caption{Curve fitting vs. nearest neighbor regression: {\tt 2PathGenerator} (top) and {\tt TriangleCount} (bottom) on {\tt web-Stanford}.}
\label{fi:curve-fitting}
\end{figure}

As a final experiment, we analyzed the interplay between nearest neighbor regression and curve fitting. Using Equation~\ref{eq:approx-remaining-time} on the longest task, we computed the percentage of $\widetilde{r}_i(t)$ obtained using curve fitting, i.e., using the $\widetilde{f}$ defined in Equation~\ref{eq:curve-fitting}. In {\tt 2PathGenerator}, more than $75\%$ of the predicted time is always due to curve fitting (Figure~\ref{fi:curve-fitting}a), while the percentage is negligible (even zero) for {\tt TriangleCount} (Figure~\ref{fi:curve-fitting}c), where all keys are very small and similar.

We then evaluated the accuracy of a variant of {\tt NearestFit} that is restricted  to use only curve fitting (called {\tt Fit} in Figure~\ref{fi:curve-fitting}). As expected, whenever there is no significant skewness and reduce executions are all very short, the accuracy of {\tt Fit} can be very poor, as in Figure~\ref{fi:curve-fitting}d. Hence, the orderly combination of the two techniques appears to be crucial to obtain good progress estimates.

\section{Related work}
\label{se:related-work}

Designing profiling methodologies able to provide insights into the performance of data-parallel jobs and their scaling properties is extremely difficult. In this section we focus on the class of MapReduce computations, which are the target of this paper. 

HiTune~\cite{HiTune} is a lightweight performance analyzer for Hadoop that allows users to identify application hot-spots and hardware problems. PerfXplain~\cite{perfXplain12} addresses performance debugging by comparing the profiles of different MapReduce jobs: programmers specify performance queries and the tool generates explanations obtained from the analysis of a log of past job executions. Similarly, self-tuning systems such as Starfish~\cite{starfish11} can help users understand and optimize many configuration parameters of Hadoop. With respect to our work, these papers target different goals.


A different line of research has addressed the problem of performance prediction in the context of MapReduce-style applications. Accurate predictions can be exploited for different practical purposes, ranging from more precise progress indicators to different optimization strategies able to reduce the job completion time.
Parallax~\cite{parallax} is a progress indicator for MapReduce pipelines based on a linear prediction model similar to {\tt JobRatio}. It can use job profiles collected on past executions. Paratimer~\cite{paratimer} extends Parallax towards workflows characterized by complex DAGs: it can predict the progress of parallel queries by analyzing the critical path in the computation. 
ARIA~\cite{ARIA11} is a framework for automatically allocating the proper amount of cluster resources in order to complete a job within a certain (soft) time deadline. Exploiting job profiles collected on past executions, the linear performance model proposed by ARIA determines the task parallelism sufficient to meet the deadline. To the best of our knowledge (and as shown in this paper), none of these  works is able to  predict accurately the running time of MapReduce applications in presence of both  data skewness and super-linear reduce functions.


Uneven data distribution is one of the main reasons for struggler tasks in MapReduce jobs. Several works~\cite{skewtune, Le2014, chisel, Gufler2011} have thus approached this challenge: their main idea is to detect stragglers and split them as soon as possible in order to reduce the job completion time. Akin to this paper, \cite{Gufler2011} considers the problem of super-linear reduce implementations alongside data skewness, but requires a user-defined cost model for predicting the reduce running time. Moreover, the size $|V_k|$ of the key group processed by a reduce function is not collected explicitly, but is estimated as an average among the sizes of all the key groups assigned to the reduce task. Though efficient, this could be rather inaccurate. Load balancing for reduce tasks is also addressed in~\cite{Le2014}, where map outputs are sampled and  the most frequent keys are detected: this is similar to our notion of explicit keys, but we avoid sampling thanks to the use of streaming data structures.

\section{Concluding remarks}
\label{se:conclusions}

In this paper we have introduced the {\tt NearestFit} progress indicator. {\tt NearestFit} targets accuracy of progress predictions, even in the presence of data skewness and super-linear reduce implementations, thanks to a careful combination of two learning techniques: nearest neighbor regression and statistical curve fitting. It also targets practical feasibility by exploiting different space-efficient data structures and data streaming algorithms. An extensive empirical assessment over the Amazon EC2 platform and several benchmarks has confirmed the precision and effectiveness of our model, that can be accurate even where competitors can be seriously harmed (i.e., in the presence of high computation times and skewed data). We believe that the better accuracy of {\tt NearestFit} can be beneficial also in other settings, where performance prediction is used as a building block for pursuing sophisticated profile-guided optimizations.


There are some assumptions that can be regarded as threats to validity of our approach:
\begin{itemize}

    \item {\em Performance as a function of the input size}. Our model makes the assumption that $f(k_1,V_{k_1})\approx f(k_2,V_{k_2})$ whenever $|V_{k_1}|\approx |V_{k_2}|$. In some applications, the performance could instead depend on the input values, and not on the input size: different values might yield quite different processing times even for key groups with similar size. We are not aware of MapReduce applications where this happens, and in general our approach is akin to algorithmic asymptotic analysis and to previous input-sensitive profiling works~\cite{CDF12,GAW07}. 


    \item {\em Uninformative profiling data}. Applications characterized by a very small number of distinct reduce keys can undermine our approach, since the profiling data collected across all tasks may be uninformative, preventing {\tt NearestFit} to take benefit of both nearest neighbor regression and curve fitting. These applications, however, would not exploit parallelism at all in the reduce phase, and are unlikely to happen in realistic big data computing scenarios. 
  
\end{itemize}

\noindent Two aspects that we have not addressed explicitly in this paper are task failures and  heterogeneous clusters.  Different progress predictions (e.g., best case vs. worst case) could be returned to assess the impact of different failure scenarios on the job progress.
%
%
With respect to heterogeneous clusters, a guiding design principle of {\tt NearestFit} is to exploit as much as possible task-specific information (points 1 and 2 in the algorithm combination described in Section~\ref{ss:nearestfit}). Global information  (points 3 and 4) is used as failsafe strategy and should take into account different node characteristics in heterogeneous clusters. As a future work, we plan to extend our implementation towards these directions.

A very interesting research problem is how to apply non-linear performance models to predict the overall progress of pipelines of jobs and workflows characterized by complex DAGs (such as those produced by Pig~\cite{pig08}). This appears to be challenging if profiles have to be dynamically collected during the actual workflow execution, without resorting to historical information.

\acks
This work is supported by Amazon Web Services through an AWS in Education Grant Award.

\bibliographystyle{abbrvnat}
\bibliography{DynamicAnalysis}

\newpage

\begin{figure*}[t]
\centering
\begin{minipage}[c]{0.52\textwidth}
\begin{center}
{\large\bf Additional examples of progress and swimlanes plots}\vspace{5mm}\\
\begin{tabular}{cc}
\hspace{-3mm}\includegraphics[width=0.50\columnwidth]{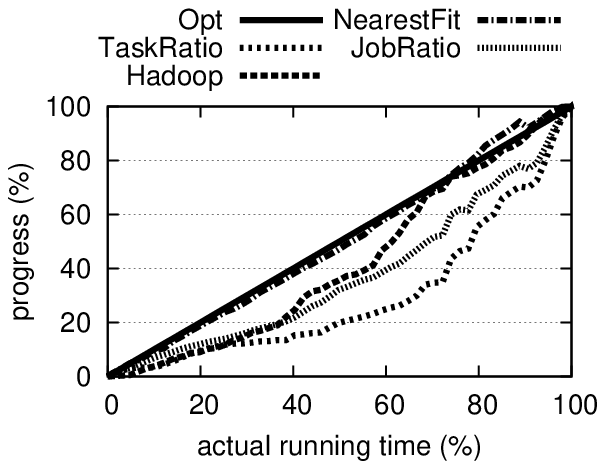} & 
\hspace{-3mm}\includegraphics[width=0.50\columnwidth]{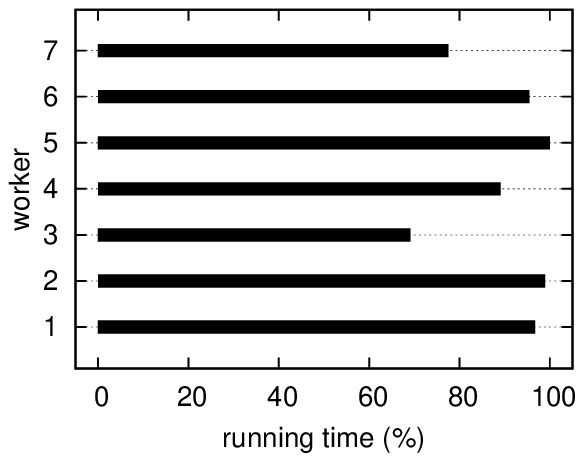} \\
\end{tabular}
\end{center}
\vspace{-2mm}
\caption{Progress and swimlanes plots: {\tt InvertedIndex} on {\tt wiki\,50K}, 8 workers, single wave.}
\label{fi:ii-50K-red7}
\end{minipage}
\vspace{4mm}
\end{figure*}

\begin{figure*}
\centering
\begin{minipage}[c]{0.52\textwidth}
\begin{center}
\includegraphics[width=0.99\columnwidth]{charts/plots/legend}\\
\begin{tabular}{cc}
\hspace{-3mm}\includegraphics[width=0.50\columnwidth]{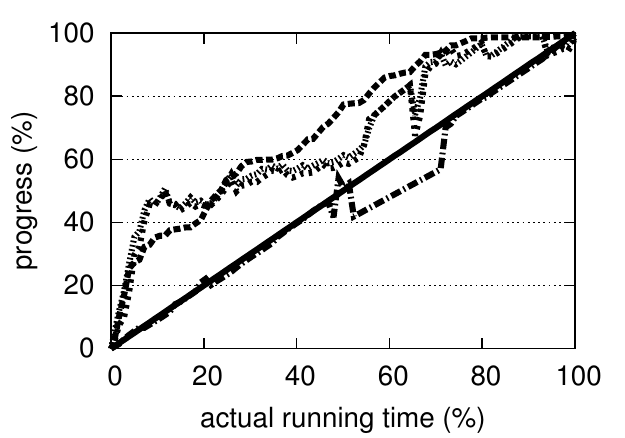} &
\hspace{-3mm}\includegraphics[width=0.50\columnwidth]{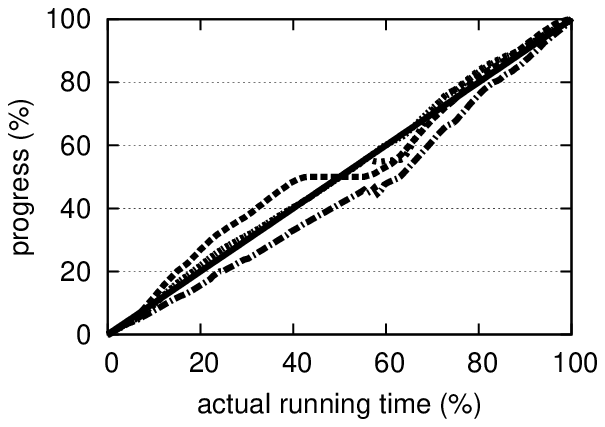} \\
\hspace{-3mm}\includegraphics[width=0.50\columnwidth]{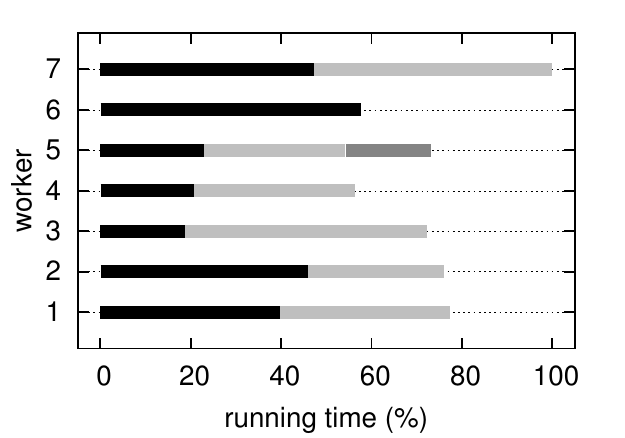} &\hspace{-3mm}\includegraphics[width=0.50\columnwidth]{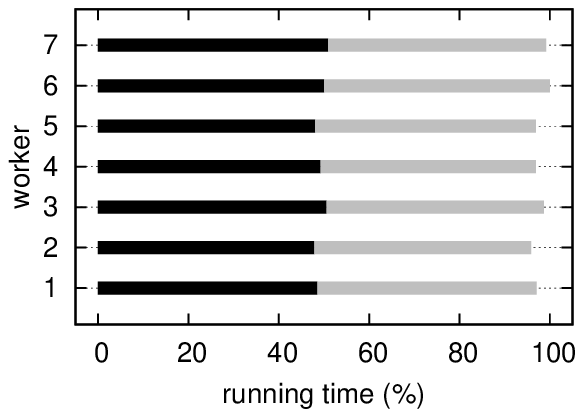}  \\
\end{tabular}
\end{center}
\vspace{-2mm}
\caption{Progress and swimlanes plots: {\tt 2PathGenerator} (left) vs. {\tt TriangleCount} (right) on {\tt as-Skitter}, 8 workers, double wave.}
\label{fi:linear-quadratic-red14}
\end{minipage}
\vspace{4mm}
\end{figure*}

\begin{figure*}
\centering
\begin{minipage}[c]{0.52\textwidth}
\begin{center}
\includegraphics[width=0.99\columnwidth]{charts/plots/legend}\\
\begin{tabular}{cc}
\hspace{-3mm}\includegraphics[width=0.50\columnwidth]{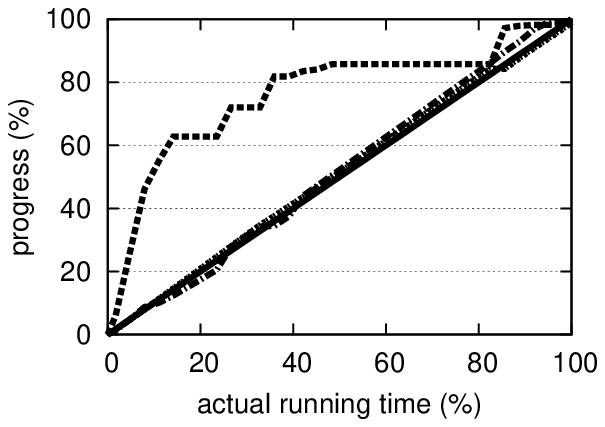} &
\hspace{-3mm}\includegraphics[width=0.50\columnwidth]{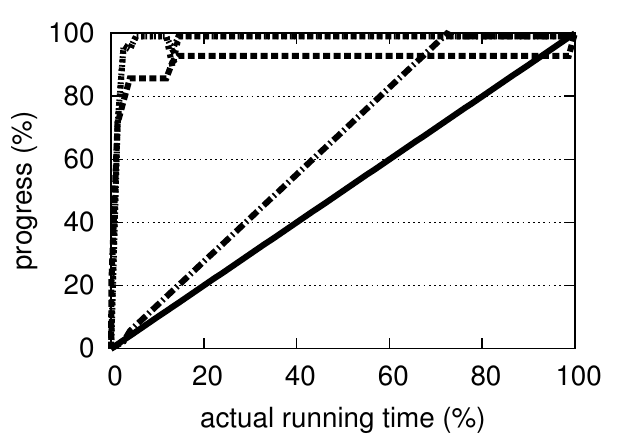} \\
\hspace{-3mm}\includegraphics[width=0.50\columnwidth]{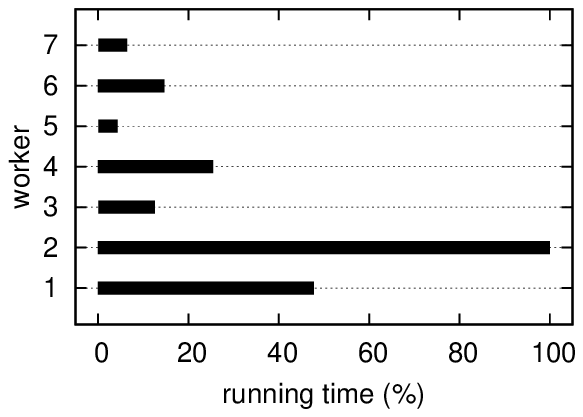} &
\hspace{-3mm}\includegraphics[width=0.50\columnwidth]{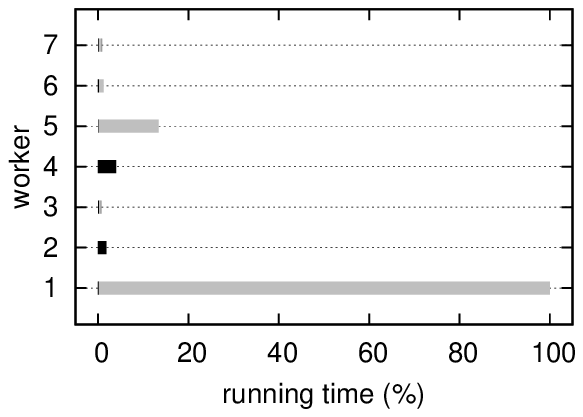} \\
\end{tabular}
\end{center}
\vspace{-2mm}
\caption{
Progress and swimlanes plots of {\tt NaturalJoin}, 8 workers, on dataset {\tt linear\,1.5} with single wave (left) and dataset {\tt sl\,1.5} with double wave (right).}
\label{fi:nj}
\end{minipage}
\end{figure*}

\end{document}